\documentclass[12pt,reqno]{amsart}
\usepackage{graphicx,amsmath,amsfonts,amssymb,bbm,color}
\usepackage{psfrag}
\usepackage{epsf}
\usepackage{latexsym}
\usepackage{rotating}
\usepackage{soul}

\DeclareMathOperator{\sech}{sech}

\begin{document}
	
\title[Synchronous collisions of KdV solitons]{Properties of synchronous collisions of solitons \\ in the Korteweg -- de Vries equation}
%\title[Statistics of synchronous collisions of KdV solitons]{Statistical properties of synchronous collisions \\ of solitons in the Korteweg -- de Vries equation}
	
	\author[T.V. Tarasova]{Tatiana V. Tarasova}
\address[T.V. Tarasova]{National Research University – Higher School of Economics, Nizhny Novgorod, Russia and Institute of Applied Physics, Nizhny Novgorod, Russia} 
	
	\author[A.V. Slunyaev]{Alexey V. Slunyaev}
	\address[A.V. Slunyaev]{National Research University – Higher School of Economics, Nizhny Novgorod, Russia and Institute of Applied Physics, Nizhny Novgorod, Russia}

\begin{abstract}
Synchronous collisions of solitons of the Korteweg -- de Vries equation are considered as a representative example of the interaction of a large number of solitons in a soliton gas.
Statistical properties of the soliton field are examined for a model distribution of soliton amplitudes according to a power law. $N$-soliton solutions ($N \le 50$) are constructed with the help of a numerical procedure using the Darboux transformation and 100-digits arithmetic.
It is shown that there exist qualitatively different patterns of evolving multisoliton solutions depending on the amplitude distribution.
Collisions of a large number of solitons lead to the decrease of values of statistical moments (the orders from 3 to 7 have been considered). The statistical moments are shown to exhibit long intervals of quasi-stationary behavior in the case of a sufficiently large number of interacting solitons with close amplitudes. These intervals can be characterized by the maximum value of the soliton gas density and by “smoothing” of the wave fields in integral sense. The analytical estimates describing these degenerate states of interacting solitons are obtained.
\end{abstract}

\date{\today}
\maketitle

\section{Introduction}
\label{sec:0}

In physics solitons represent isolated waves, maintained by a stable balance between the counteracting effects of nonlinearity and dispersion. They exhibit an exceptional stability which becomes apparent through elastic interactions with other solitons and quasilinear waves. They also demonstrate specific dynamical properties, such as velocities which differ from the ones of linear waves; peculiar scenarios of pairwise collisions; specific characteristic scales of response to a slow change in propagation conditions or weak external forcing, etc. Waves of soliton type are diverse (in the form of bell-shaped perturbations, wave groups, kinks, breathers, and so on) and are observed in various physical realms (hydrodynamics, nonlinear optics, plasma, electronics, etc.). 
The relevant literature is extensive and continues to grow.

The specificity of soliton dynamics is also manifested in probabilistic properties. While fields of completely incoherent random waves by virtue of the Central Limit Theorem are described by the Gaussian probability distribution, coherent wave interactions due to the presence of solitons can lead to essentially non-Gaussian statistics \cite{Onoratoetal2016,Randouxetal2016}, which nowadays has practically no theoretical description.
Solution of such a problem is among the main goals of the research of so-called ``rogue waves’’ in the ocean. There, the neglect of the effect of wave self-modulation due to the 4-wave quasi-resonant interactions leads to a noticeable underestimation of the probability of occurrence of high-amplitude waves \cite{Kharifetal2009,Onoratoetal2013,avs}.
Similar effects of rogue waves, associated with nonlinear instabilities and emergence of soliton-like long-lived structures were observed in a number of other physical applications \cite{AkhmedievPelinovsky2010,Onoratoetal2013,Dudleyetal2019,Rozentaletal2021}.
The limiting case of predominance of solitons in a wave field occupies a highly important place in the problems connected with the signal transmission in optical communication lines. For them\, the probabilistic description of the electromagnetic wave intensity and the control of probability of extreme field occurrence (and subsequent electrical breakdown) are also important and relevant \cite{Sollietal2007,Soto-Crespoetal2016}.  

The key role of solitons in solving equations of mathematical physics became clear with the invention of the Inverse Scattering Transform (IST) \cite{Novikovetal1984,absigur}, which provides a principle way to integrate a class of nonlinear partial differential equations. In particular, this method enables one to obtain analytical solutions. From this perspective, solitons correspond to waves of the discrete spectrum of the associated scattering problem, which do not disperse and represent the large-time asymptotic solution of the Cauchy problem for localized initial conditions. Note that the solution of the inverse scattering problem (i.e., reconstruction of waves from the spectral data) turns out to be much easier for the soliton part of solution.

Thus, in a number of cases a soliton ensemble or a \textit{soliton gas} is an appropriate and convenient approximation of a complex nonlinear wave field. Within the framework of integrable equations, the evolution of each realization of a soliton gas corresponds to a completely deterministic dynamics. However, it is so complicated that can be considered as chaotic (so-called ``soliton turbulence”) and then requires probabilistic approaches to the description. Initial ``phases'' of solitons (which can be understood as positions of solitons, complex phases of envelope solitons, etc.) play the role of random parameters of the soliton gas together with soliton amplitudes, which may be chosen in each realization according to some probability distribution.

Kinetic equations were derived to describe the soliton gas dynamics \cite{zakharov,elkamchatnov}. They characterize the transport of the soliton spectral density, but due to the violation of the wave linear superposition principle, do not provide information about the wave solution itself (which can be water surface displacement, intensity of electromagnetic fields, etc.). In particular, the questions about the probability distribution for wave amplitudes or  about the values of the wave field statistical moments remain unanswered. Multisoliton solutions, which can be formally written in a closed form using the IST or related methods for integrable equations, are very cumbersome, what makes their analytical and even numerical analysis difficult. The direct numerical simulation of evolution equations is commonly used to study the soliton gas evolution, which also becomes complicated in the case of a dense gas (i.e. when many solitons interact simultaneously), see examples in \cite{Didenkulovaetal2019}.

The consideration of simplified toy problems seems to be a reasonable approach to understanding the laws of soliton gas dynamics. Probabilistic characteristics of pairs of interacting solitons were analyzed in \cite{Pelinovskyetal2013}, where the pairwise collisions were considered as ``elementary acts'' of the soliton turbulence. This idea has confirmed its fruitfulness in a series of subsequent publications (\cite{Shurgalina2018,PelinovskyShurgalina2017} and others) in a number of examples within the framework of equations of the Korteweg -- de Vries type. 
``Synchronous'' collisions of solitons \cite{SlunyaevPelinovsky2016,Sun2016,Slunyaev2019} is another more complicated representative case. 
The limiting case of collision of an infinitely large number of solitons was studied in the papers \cite{BilmanBuckingham2019,BilmanMiller2021} within the framework of the focusing nonlinear Schr\"odinger equation. 

In the present paper synchronous interactions of a large number of solitons of the Korteweg -- de Vries (KdV) equation are investigated. All the solitons are characterized by relative positions, which correspond to the origin, $x=0$, at the moment of the ``focusing” $t=0$. Though the $N$-soliton expressions for the KdV equation can be formally written explicitly, they are too complicated for analytical analysis and are studied by numerical methods. The numerical construction of the $N$-soliton solution is non-trivial when $N$ is not small ($N\gtrsim5$). In this paper we develop and apply a method for constructing multisoliton solutions, which is based on the Darboux transformation and uses the representation of numbers with a long mantissa, as proposed in \cite{GelashAgafontsev2018} for the case of the nonlinear Schr\"odinger equation. Our code returns highly accurate solutions for $N \lesssim 50$. 

The paper is organized as follows. The method and the approach to controlling the accuracy of the obtained solutions are described in Sec.~\ref{sec:ExactSolutions}.  
The implemented numerical procedure is used in Sec.~\ref{sec:Colliisons} for the investigation of patterns of evolving multisoliton wave fields of the KdV equation with a model distribution of  soliton amplitudes. 
%The time dependencies of instantaneous values of the statistical moments are constructed for different numbers of synchronously interacting solitons.
%
%Scenarios of soliton collisions, implementing the limiting case of the reduction of the statistical moments, are demonstrated.  It is discussed that they also correspond to situations of a critical density of a soliton gas.
%
The results of the study are summarized in the final Sec.~\ref{sec:Conclusion}.  

\section{Construction of exact solutions of the KdV equation with large number of solitons} \label{sec:ExactSolutions}

In the present paper the standard dimensionless form of the KdV equation
\begin{align}\label{kdv}
	u_t + 6 u u_{x} + u_{xxx} = 0
\end{align}
is used for the real function $u(x, t)$. The variable $x \in (-\infty, +\infty)$ serves as a space coordinate, $t \in (-\infty, +\infty)$ is the time.

There is a number of methods of constructing multisoliton solutions of integrable equations. Several algorithms were implemented and examined at the preliminary stage of this study. One of them was based on the solution of the Gel'fand -- Levitan -- Marchenko equation for the inverse scattering problem \cite{absigur}, other were implemented with the use of the Hirota $\tau$-function \cite{newell} and with the help of the Darboux transformation \cite{matveevsalle}. The procedure of numerical construction of solutions using the Darboux transformation appeared to be the most successful compared to other approaches in terms of the computation speed and accuracy of the solution. 

In this section we provide the mathematical formulation of the Darboux transformation for the KdV equation, the features of its implementation for numerical construction of $N$-soliton solutions with a large number of solitons, and the method for controlling the accuracy of the obtained solutions.
%along with its results.

\subsection{The Darboux transformation for synchronous collisions of solitons}

It is a well-known fact that  (\ref{kdv}) is a compatibility condition for the system of equations called the Lax pair \cite{Novikovetal1984,absigur}:
\begin{align}\label{Lax}
	\left\{
	\begin{aligned}
	\hat{L} \Psi = \lambda \Psi \\
	\hat{A} \Psi = \Psi_t,
\end{aligned} \right.  
\end{align}
where the operators $\hat{L}$ and $\hat{A}$ are introduced as follows: 
\begin{align} \label{LA}
	\hat{L}  = -\frac{\partial^2}{\partial x^2} - u(x,t), \qquad
	\hat{A}  = -4\frac{\partial^3}{\partial x^3} - 6u\frac{\partial}{\partial x} - 3u_x.
\end{align}
In (\ref{Lax}) the real constants $\lambda$ are eigenvalues of the scattering problem (\ref{Lax}a) (a spectrum of the scattering problem). Negative values of $\lambda$ correspond to the discrete spectrum. One eigenvalue $\lambda = - k_1^2$ specifies one soliton of the KdV equation in the form
\begin{align} \label{OneSoliton}
	u_1(x,t) = 2k^{2}_1 \sech^2 \left(k_1(x-4k^2_1t-x_1) \right).
\end{align}
Here the parameter $k_1$ determines the amplitude $A_1 = 2k_1^2>0$ and the velocity $V_1 = 4k_1^2>0$ of the soliton. The value $x_1$ specifies a relative position of the soliton in space. Solitons of the KdV equation (\ref{kdv}) always have a positive polarity, $u_1(x,t)>0$.

The Darboux method is based on the covariance of the Lax pair with respect to the transformation
\begin{align}
	\widetilde{u}(x,t) = u + 2\frac{\partial \gamma}{\partial x}, \quad
	\widetilde{\Psi}(x,t)=\frac{\partial \Psi}{\partial x} - \gamma \Psi, \quad 
	\gamma(x,t) = \frac{\partial \Psi_1}{\partial x}\frac{1}{\Psi_1}, 
\end{align}
where $\Psi_1(x,t)$ is some particular solution to the system of equations (\ref{Lax}) when $\lambda = \lambda_1$ \cite{matveevsalle}. In the general case the function $\widetilde{u}(x,t)$ is a new non-trivial solution of (\ref{kdv}), and the function $\widetilde{\Psi}(x,t)$ is the corresponding eigenfunction of this new solution.

The $N$-fold Darboux transformation with the use of $N$ “seed” functions $\Psi_j$, $j=1,...,N$, enables to construct the required $N$-soliton solutions to (\ref{kdv}). The result of the $N$-fold
Darboux transformation may be expressed as follows \cite{matveevsalle}:
\begin{align} \label{Darboux}
	\widetilde{u} = u + 2\frac{\partial^2}{\partial x^2} \ln W_N (\Psi_1, \Psi_2, ... \Psi_N),  \qquad
	\widetilde{\Psi} = \frac{W_{N+1}(\Psi_1, \Psi_2, ... \Psi_N, \Psi)}{W_{N}(\Psi_1, \Psi_2, ... \Psi_N)},
\end{align}
where the Wronskians are composed of derivatives of the seed functions:
\begin{align} \label{Darboux_mat}
	W_{N} =
	\begin{vmatrix}
		\Psi_1 & \dots & \Psi_N \\ 
		\hdotsfor{3} \\
		\Psi_1^{(N-1)} & \dots & \Psi_N^{(N-1)}
	\end{vmatrix}, \qquad
	W_{N+1} =
	\begin{vmatrix}
		\Psi_1 & \dots & \Psi_N & \Psi \\ 
		\hdotsfor{4} \\
		\Psi_1^{(N)} & \dots & \Psi_N^{(N)} & \Psi^{(N)}
	\end{vmatrix}.
\end{align}
Here $\Psi_j^{(k)} := \frac{\partial^k  \Psi_j}{\partial x^k}$, and $\Psi_j(x,t)$ are some particular solutions to the system of equations (\ref{Lax}) with corresponding eigenvalues $\lambda = \lambda_j$.

In order to construct an $N$-soliton solution of the KdV equation
%$\widetilde{u}(x,t)$, 
one should use the trivial initial solution, $u(x,t) \equiv 0$, and take the following functions as the particular solutions to the system of equations (\ref{Lax}) with the zero potential:
\begin{align}\label{seed}
	\Psi_{j} (x,t) = \Phi_j{(k_{j}(x - x_j) - 4k_{j}^3t)}, \qquad \text{where} \qquad \qquad \qquad\\
	\Phi_j(\cdot) \equiv \cosh{(\cdot)}, \quad \text{if $j$ is even,}
	\quad \text{and} \quad \Phi_j(\cdot) \equiv \sinh{(\cdot)}, \quad \text{if $j$ is odd}, \quad j = 1,...,N. \nonumber
\end{align}
Here $\lambda_j = - k_j^2 < 0$, and arbitrary constants $x_j$ specify the relative positions of the solitons in space. The dependence of phases of eigenfunctions on time is determined by the second equation of the Lax pair (\ref{Lax}b). 

The choice of different parameters $k_j>0$, $j=1,...,N$ and alternating seed functions (\ref{seed}) ensures the non-degeneracy of the Wronskians (\ref{Darboux_mat}) and lead to the exact solution, 
\begin{align}\label{sol}
	u_N(x,t) = 2\frac{\partial^2}{\partial x^2} \ln W_N (\Psi_1, \Psi_2, ... \Psi_N),
\end{align}
describing a non-linear superposition of $N$ solitons with amplitudes $A_j = 2k^2_j$, velocities $V_j = 4k^2_j$ and relative initial positions $x_j$. In particular, for $N=1$ one obtains one-soliton solution in the form (\ref{OneSoliton}).

Since the phases of functions $\Phi_j$ contain different velocities of solitons $V_j=4k_j^2$, for large times $|t| \gg 1$ the solution (\ref{sol}) asymptotically tends to a linear superposition of solitons which are located far apart from each other:
\begin{align} \label{Asymptotics}
	\lim_{t \to \pm\infty} u_N(x,t) = \sum_{j=1}^{N} 2k_j^2 \sech^2 \left(k_j(x-4k_j^2t-x_j\pm\delta_j) \right).
\end{align}
At large times the solitons are ordered according to their velocities. In (\ref{Asymptotics}) the parameter $\delta_j$ corresponds to the shift of coordinate of the $j$-th soliton, obtained as a result of interaction with the other solitons.

It should be mentioned that the ``initial'' coordinates $x_j$ in the obtained solution are virtual. In the general case, at the moment of time $t = 0$ the maximum of the $j$-th soliton may be located at a point different from $x = x_j$. Meanwhile, a particular choice of zero values for these parameters, $x_j=0$, $j=1,...,N$, leads to the following symmetry of the solution: $u_N(x,t)=u_N(-x,-t)$ 
for any given $N$.
In particular, the solution is even with respect to the origin, $u_N(x,0)=u_N(-x,0)$ and $u_N(0,t)=u_N(0,-t)$; the function of maximum of the solution, $\max_x {u_N}(t)$, is symmetric with respect to $t=0$. Thus, the choice of zero values for all $x_j$
%, $j=1,...,N$, 
ensures the synchronous interaction of solitons, what should correspond to the strongest focusing of solitons at $t=0$.

\subsection{Numerical procedure for exact $N$-soliton solutions}

\begin{figure}[htp]
	{\includegraphics[width=7cm]{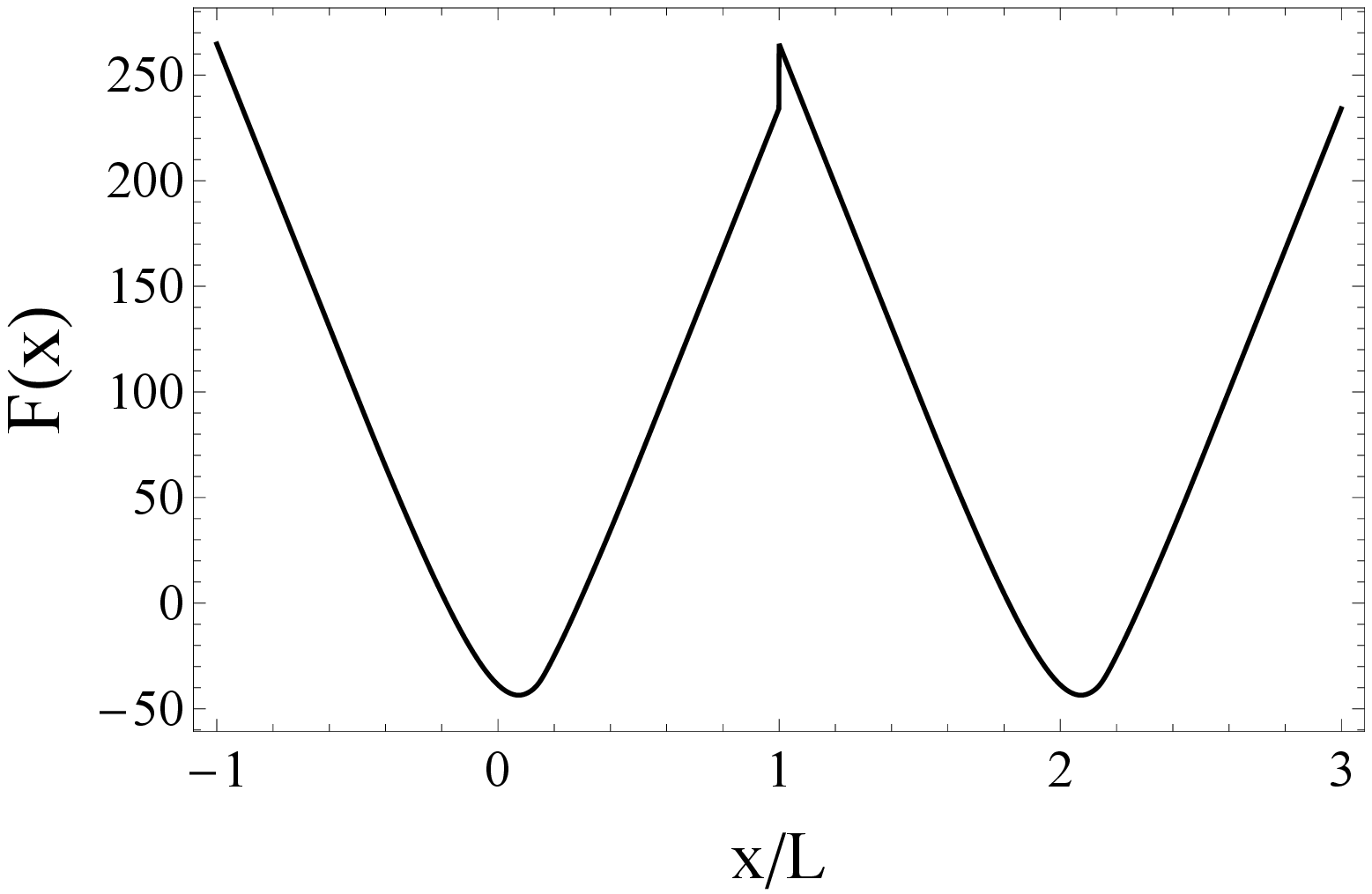}}(a)
	{\includegraphics[width=7cm]{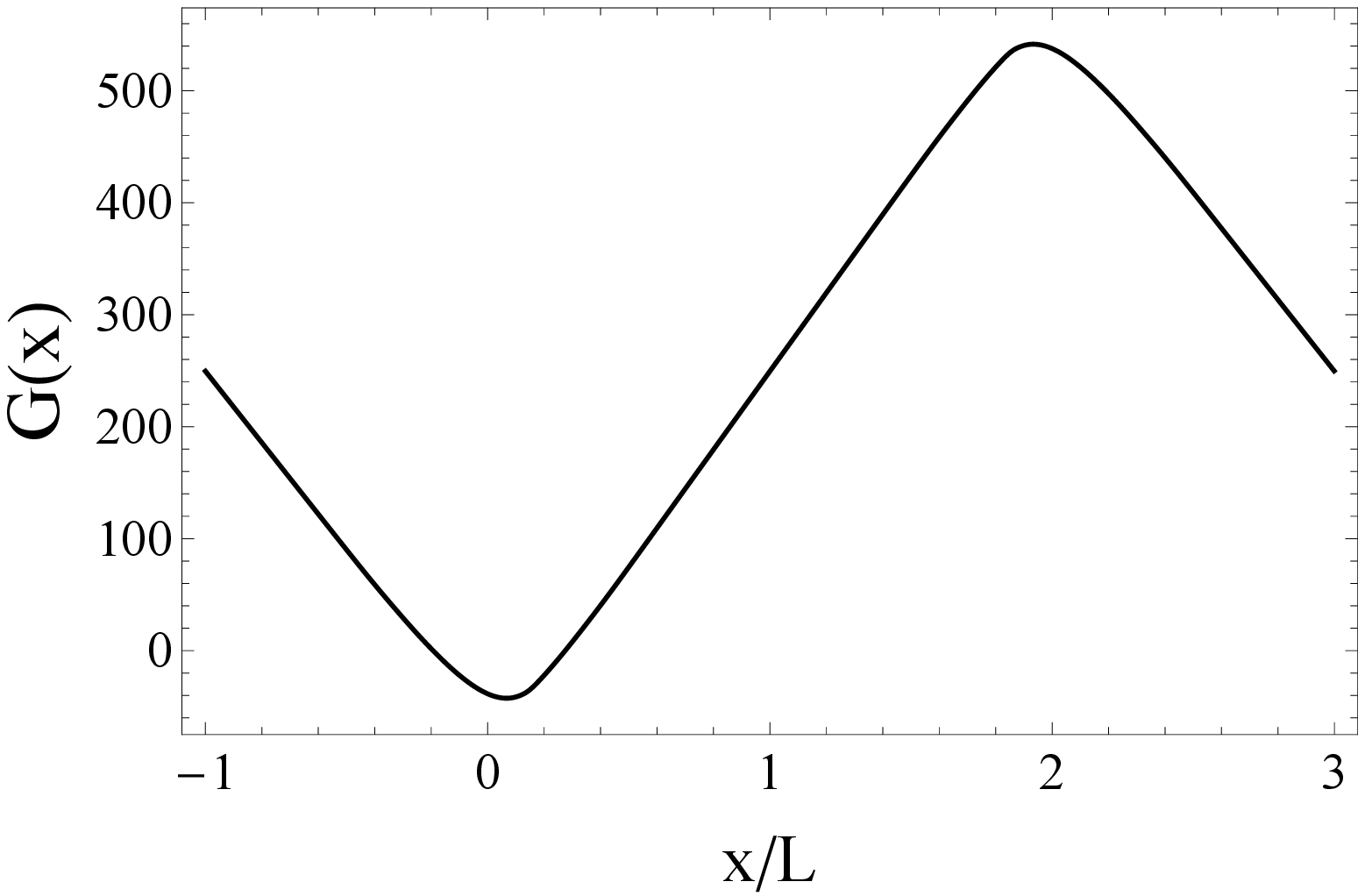}}(b)
	\caption{An example of the auxiliary functions for the $N$-soliton solution: a)~the periodic extension $F(x)$ of the function $f(x) = \ln W_N(x)$, b)~the function $G(x)$.}
	\label{fig:WronskianContinuation}
\end{figure}

The exact $N$-soliton solution, given by the formula (\ref{sol}) for the $N$-fold Darboux transformation, is calculated numerically for given instants of time $t$ in nodes of a regular grid within the interval $x\in[-L, L]$.
After calculation of the function $f(x)= \ln W_N (x)$, its second derivative is required to calculate the solution $u_N$; even higher-order derivatives (up to the tenth) should be computed to check the conserved quantities (see the next subsection). 
The discrete Fourier transform is known to be efficient for calculation of high-order derivatives. In our case, it cannot be applied directly, since the periodic extension $F(x)$ of the function $f(x)$ generally has discontinuities at the points of linkage $x=\pm L$ (Fig.~\ref{fig:WronskianContinuation}a). As a result, the high-order derivatives of $f(x)$ calculated using the Fourier series exhibit considerable oscillations at the ends of the domain (the Gibbs phenomenon).

For regularization of the solution we equalize the values of the function $f(x)$ at the boundaries of the considered interval using a linear additive, which provides the new function $g(x)$, $g(-L)= g(L)$:
\begin{align}\label{z}
	g(x) = f(x) + \frac{f(-L) - f(L)}{2 L} x.
\end{align}
Then, the function $g(x)$ is extended outside the interval $[-L,L]$ in an antisymmetric way. This enables to obtain an everywhere differentiable function $G(x)$, which is periodic in the interval $[-L,3L]$ (see Fig.~\ref{fig:WronskianContinuation}b):
\begin{align}\label{G(x)}
	\left\{
	\begin{aligned}
		&G(x) = g(x), \quad -L \leq x \leq L, \\
		&G(2L+x) = 2g(L) - g(-x),  \quad -L \leq x \leq L.
	\end{aligned} \right.  
\end{align}
The multisoliton solution is constructed with the use of the modified function, $u_N = 2 \frac{\partial^2}{\partial x^2} f(x) = 2 \frac{\partial^2}{\partial x^2} G(x)$ for $x \in [-L,L]$. 

Though the formula (\ref{sol}) represents an explicit analytical expression for $u_N$, it returns inaccurate solutions for $N$ greater than $10$ or so due to the computer rounding errors. Besides, the computation time rapidly increases with $N$. The approach to calculate the Wronskian using the symbolic mathematics is time consuming and does not solve the problem of accuracy of the solutions, since a great number of terms of very different magnitudes must be summed to obtain eventually the numerical values.

To address this problem, we followed the suggestion by \cite{GelashAgafontsev2018, gelash} and used a 100-digits arithmetic during the calculations. Furthermore, in the expression of the Wronskian (\ref{Darboux_mat}) the derivatives of the seed functions were precomputed analytically; all other calculations were performed numerically with the long mantissa. The implemented procedure enables to construct solutions, consisting of a large number of solitons (up to about 50), with high accuracy.

\subsection{Accuracy control of the constructed solutions}

As discussed above, the calculated numerically exact solutions for a large $N$ may suffer from a considerable error related to the calculation of Wronskian, which is unacceptable for the purposes of the present study.
In order to estimate the accuracy of the constructed solutions, we use the property of the KdV equation as an integrable system -- the existence of an infinite sequence of non-trivial conservation laws in the form \cite{miura}
\begin{align} \label{I_m}
	\frac{d}{dt} I_m = 0, \qquad I_{m} = \int_{-\infty}^{+\infty} T_{m}(u(x,t)) dx, \qquad m = 1,2, ... .
\end{align}
The physical integrals of motion consist of combinations of these quantities \cite{AblowitzSegur1979}; the integral $I_3$ corresponds to the Hamiltonian of the equation \cite{ZakharovFaddeev1971}. 
The list of the first 10 conservation densities $T_m(x,t)$ is provided in the Appendix.

The $N$-soliton solution of the KdV equation (\ref{sol}) in the limit $t \rightarrow \pm \infty$ tends asymptotically to a linear combination of solitons, which can be located arbitrarily far apart from each other due to the propagation with different velocities $V_j = 4k_j^2$, see Eq. (\ref{Asymptotics}). As a result, each of the conservation integrals for $u_N$ is equal to the sum of the corresponding integrals for independent solitons. Expressions for the integrals $I_m$ for one soliton were provided in the book \cite{karpman} for an arbitrary order $m$.  Then the formula for the integrals of an $N$-soliton solution follows:
\begin{align} \label{IntegralsNSoliton}
	I^{(N)}_{m} =  24^m \frac{[(m-1)!]^2}{(2m-1)!}\sum_{j=1}^{N} k_j^{(2m - 1)},
\end{align}
where $k_j = \sqrt{A_j/2}$, and $A_j$ are amplitudes of solitons, as before. The obtained $N$-soliton solutions should be characterized by the integrals (\ref{IntegralsNSoliton}) for any parameter of time $t$.

Checking the conservation of the integrals of motion of a dynamical system (mass, momentum and energy) is a common approach to assessing the accuracy of the direct numerical simulation.
Any conserved densities $T_m$ cannot be expressed through the other ones by trivial transformations, hence it is natural to expect that the more integrals coincide with their analytical values (\ref{IntegralsNSoliton}), the more accurate the calculated solution is. 
As can be seen from the formulas for $T_m$ (see the Appendix), with an increase of the order $m$ the expressions for densities $T_m$ include higher degree of nonlinearity and higher order of derivatives (for the provided 10 conservation laws -- up to tenth and eighth respectively). Thus, higher order integrals become more and more sensitive with respect to deviations from the exact solution. 
The present study of geometrical and statistical properties of multisoliton fields is based on the solutions for which the relative error of every conservation integral does not exceed 3\%.

\section{Synchronous collisions of a large number of KdV solitons} \label{sec:Colliisons}

In this section the dynamics and probabilistic properties of the $N$-soliton solutions (\ref{sol}) with zero “initial” coordinates $x_j = 0$, $j=1,...,N$ are considered. The reliability of the results is ensured by a high accuracy of the multisoliton solutions obtained using the procedure described in the previous section.

\subsection{The general picture of many-soliton interactions}

Interactions of a small number of solitons were discussed many times in the classical literature with comprehensive illustrations of the solutions (see, for example, \cite{absigur}, \cite{newell}, \cite{Lamb1980}). Plotting exact solutions consisting of a large number of solitons (say more than 10) has only become possible in recent years, see, for instance, \cite{gelash} for the nonlinear Schr\"odinger equation, the recent preprint \cite{Bonnemainetal2022} for the KdV equation, the article \cite{Slunyaev2019} for the modified and generalized KdV equations.

As mentioned above, we consider the particular choice of the relative soliton positions, which ensures a synchronous collision and a high symmetry of the solutions, that should correspond in some sense to the maximal focusing of solitons at $t=0$ at the point $x=0$.

\begin{figure}[htp]
	{\includegraphics[width=7cm]{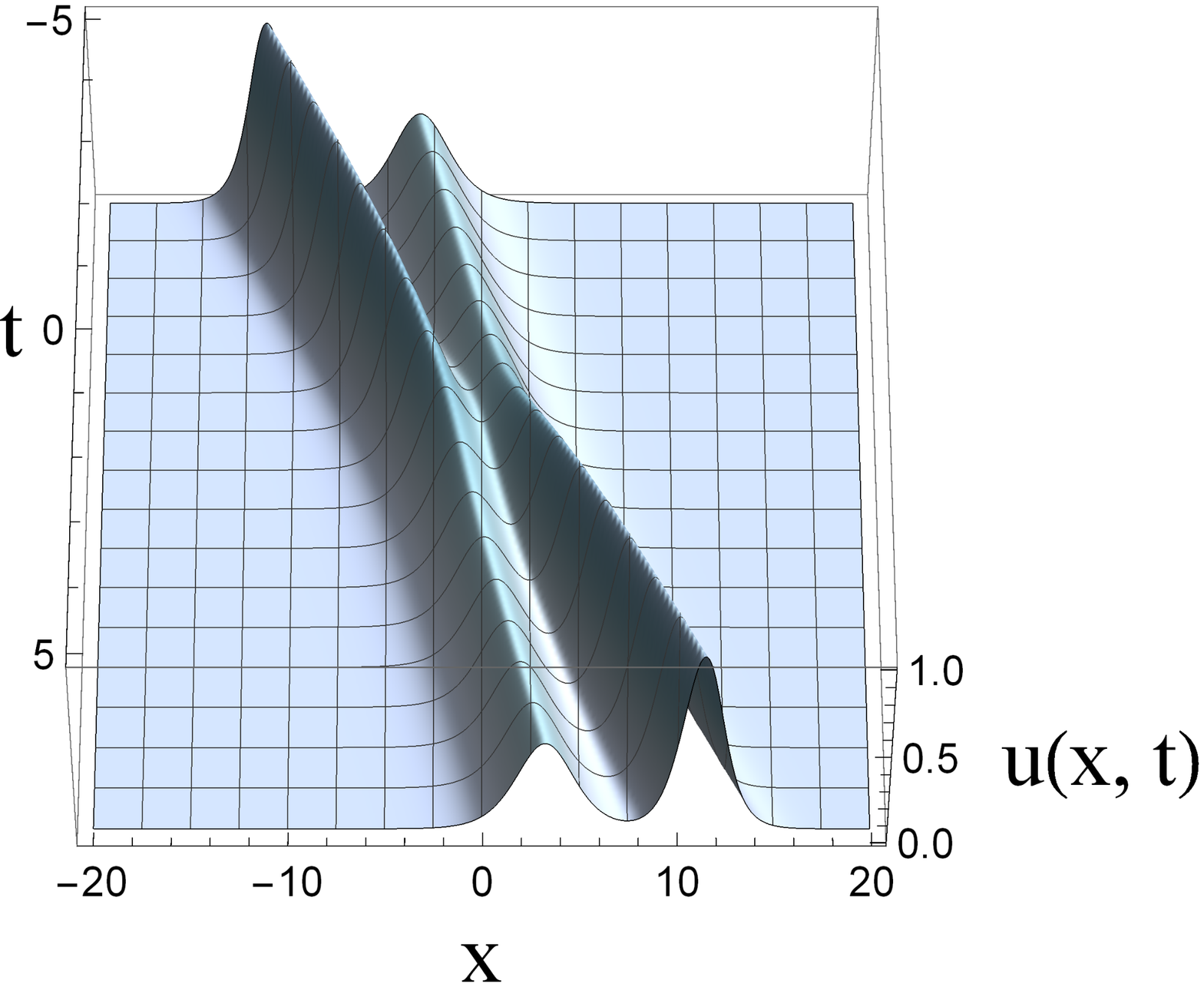}}(a)
	{\includegraphics[width=7cm]{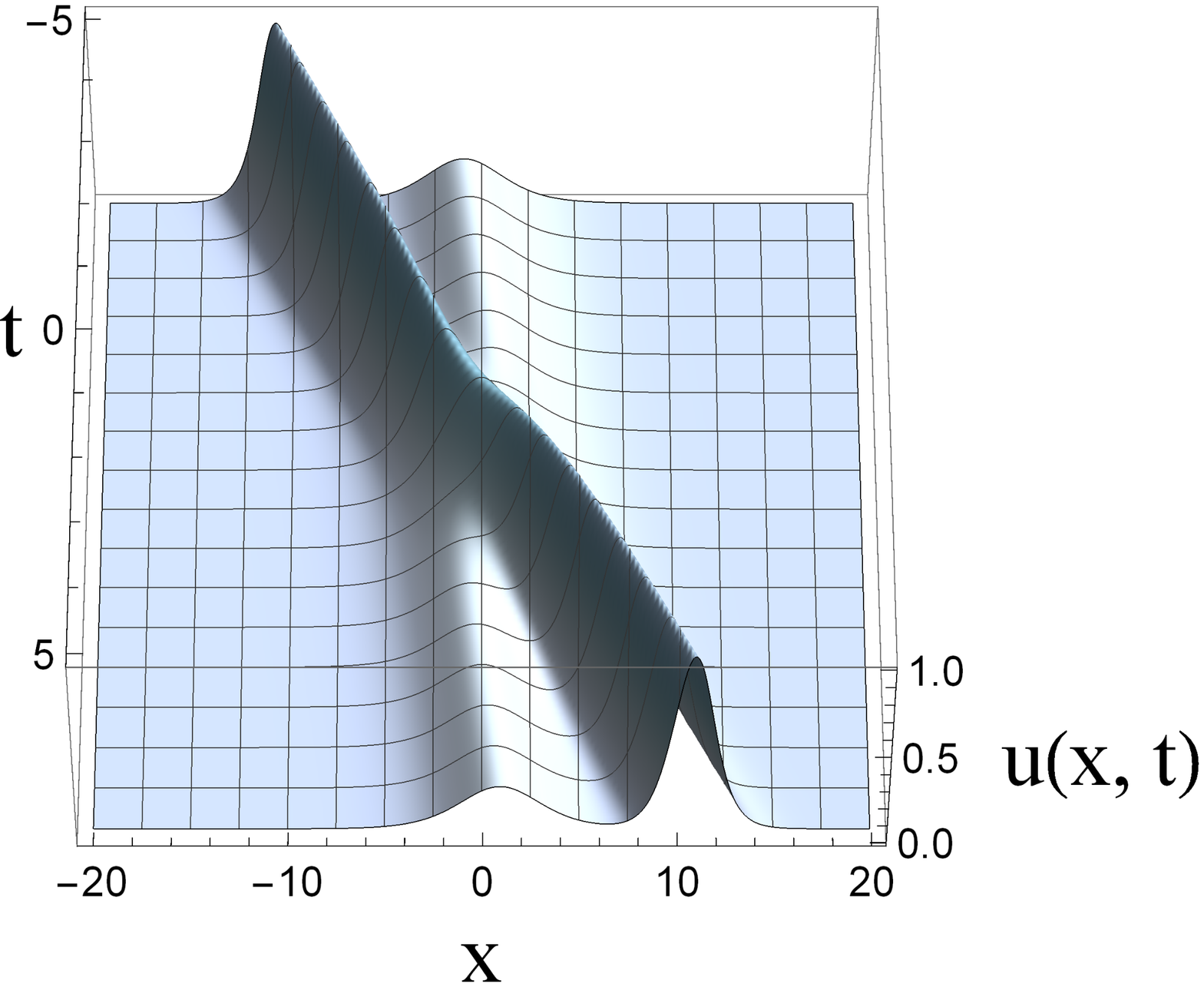}}(b)
	\caption{Exchange (a) and overtaking (b) interactions of solitons of the KdV equation.}
	\label{fig:TwoTypesOfInteracions}
\end{figure}

Let us recall that for pairs of interacting KdV solitons two regimes are commonly distinguished, which may be called ``exchange'' and ``overtaking'' collisions \cite{Lamb1980}. They differ by a two-hump (Fig.~\ref{fig:TwoTypesOfInteracions}a) or a one-hump (Fig.~\ref{fig:TwoTypesOfInteracions}b) shape of the wave at the time instant when the distance between interacting solitons is minimum. For our choice of the relative soliton positions this difference can be formalized through the sign of the second spatial derivative at the point $x = 0$ at $t = 0$, when the solution becomes an even function. It is well-known that these two types of interaction occur at different values of the amplitude ratio $d=A_1/A_2$, where for certainty $A_1>A_2$. The exchange interaction occurs between solitons with close amplitudes, $1<d<3$ (Fig.~\ref{fig:TwoTypesOfInteracions}a). Solitons with very different amplitudes, $d>3$, interact in accordance with the overtaking scenario, when the faster soliton passes through the slower one with a lower amplitude (Fig.~\ref{fig:TwoTypesOfInteracions}b). 

Hereinafter, a power-law distribution of amplitudes of $N$ solitons with the first term $A_1=1$ is considered as an example,
so that the amplitudes form a geometric progression with the common ratio $d^{-1}$:
\begin{align}\label{geom}
	A_j = \frac{1}{d^{j-1}}, \quad j = 1,2, ..., N, \quad d > 1. 
\end{align}
When $N=2$, the parameter value $d=3$ separates the two types of soliton interactions.

\begin{figure}[htp]
	{\includegraphics[width=7cm]{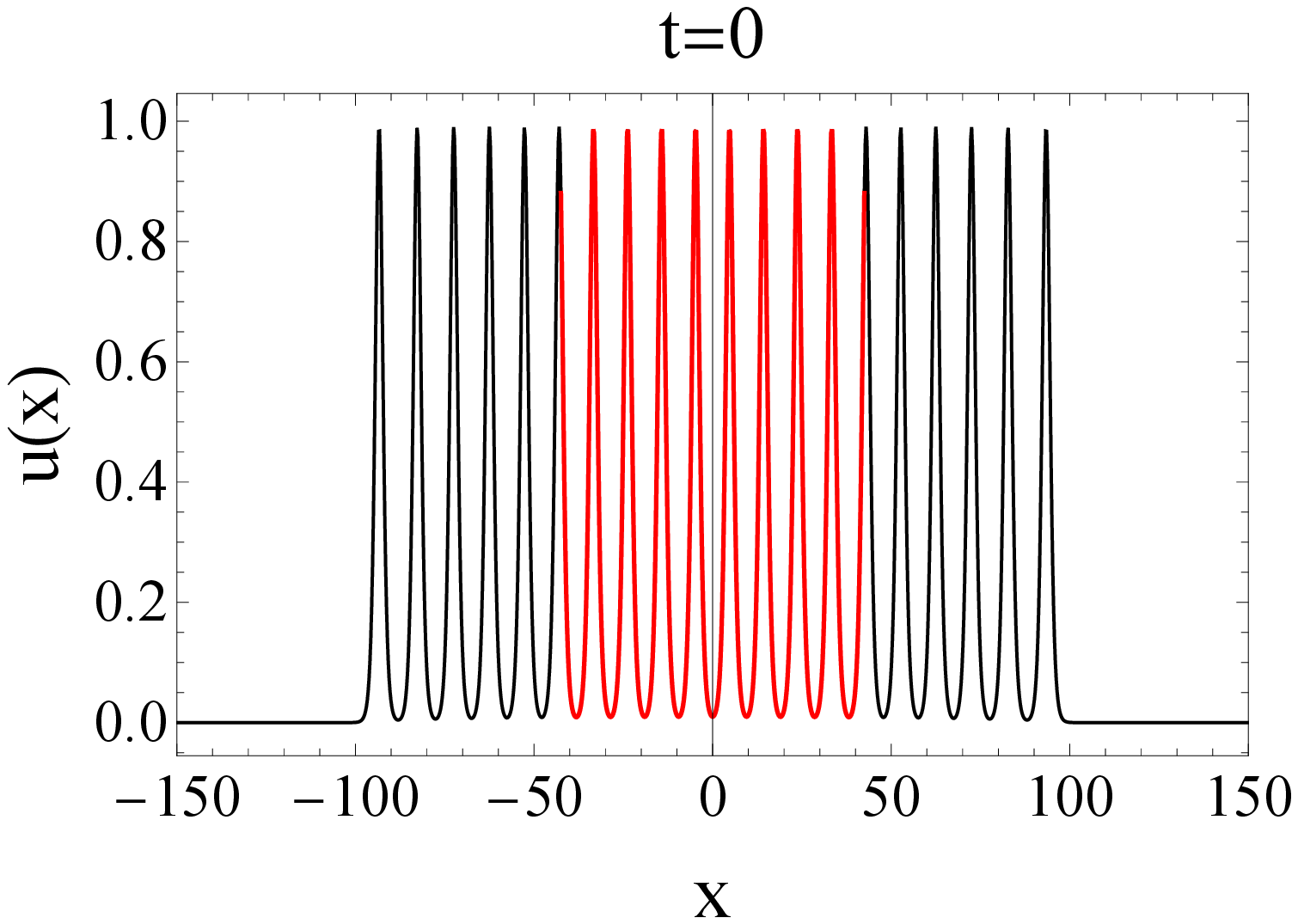}}(a)
	{\includegraphics[width=7cm]{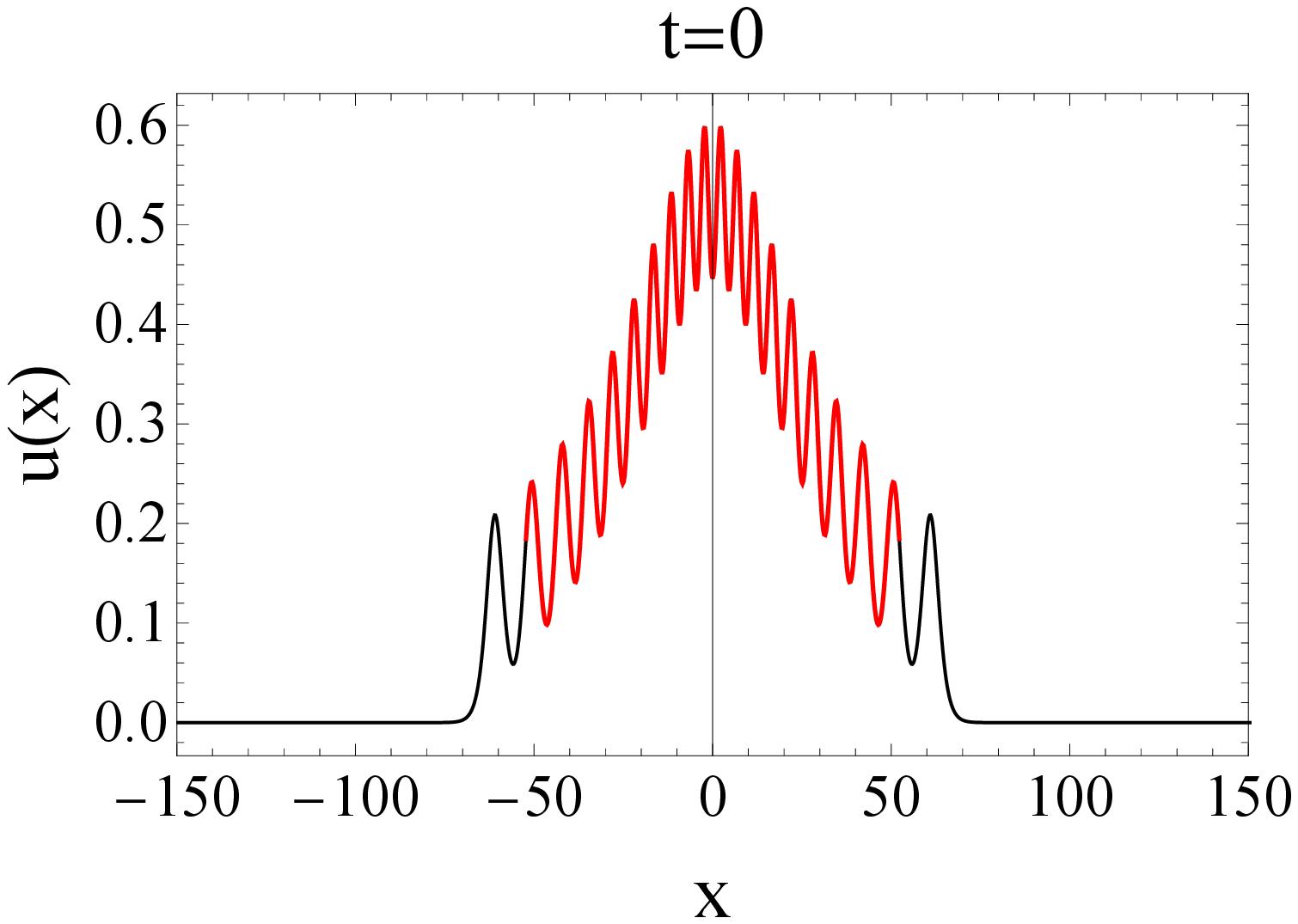}}(b)
	\\
	{\includegraphics[width=7cm]{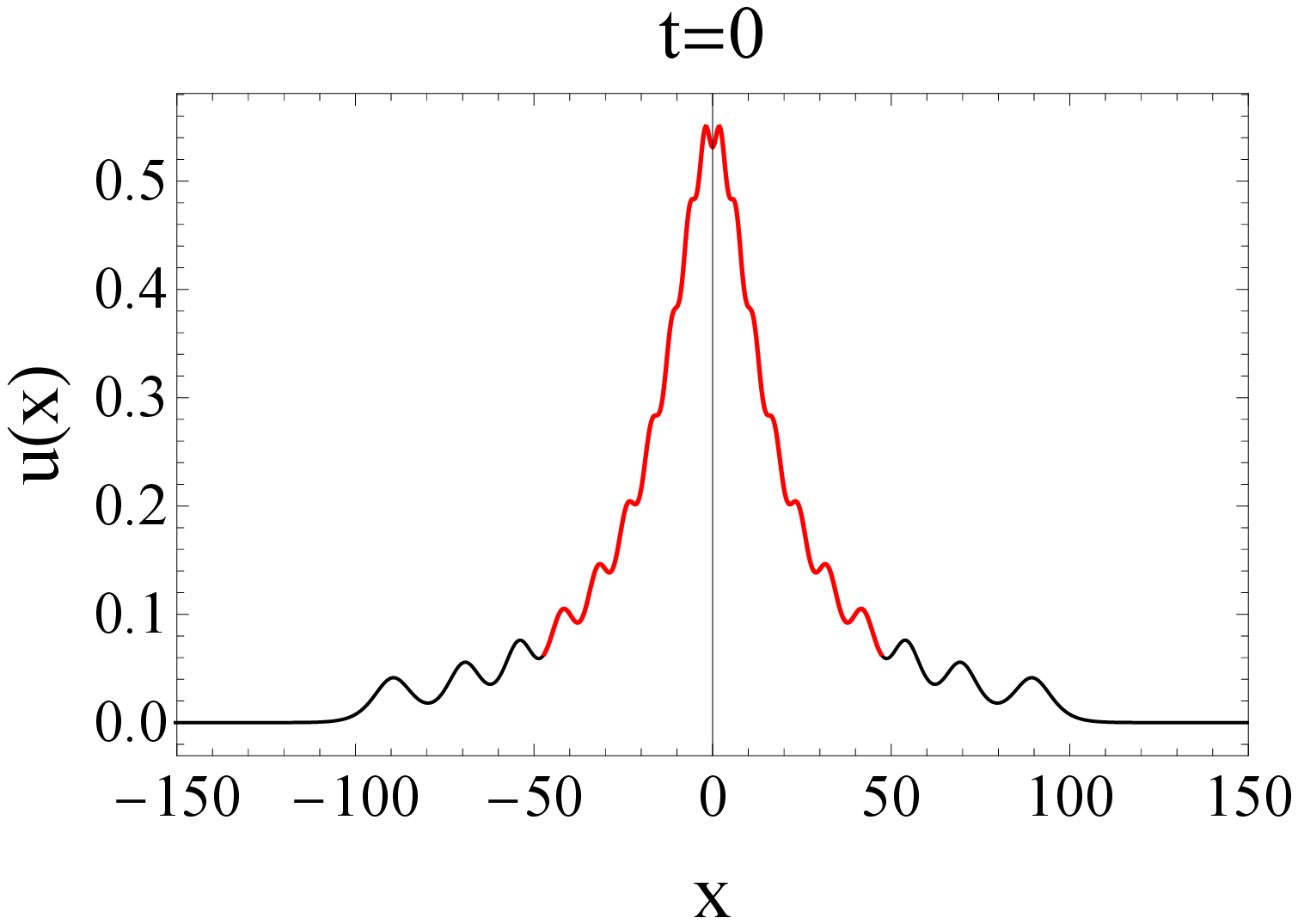}}(c)
	{\includegraphics[width=7cm]{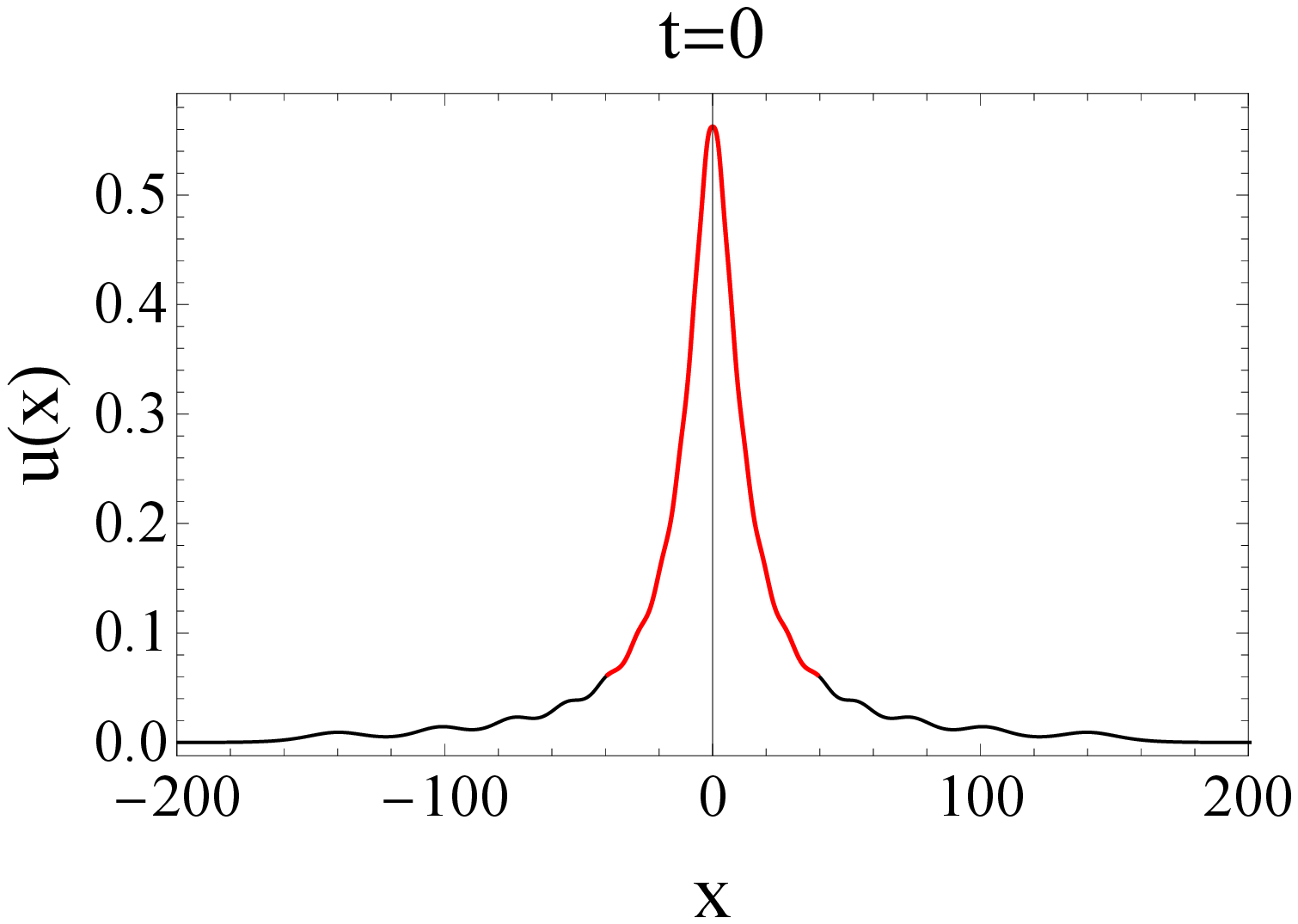}}(d)
	\caption{20-soliton solution $u_{20}(x, t)$ at the moment of time $t = 0$ for different values of $d$: a) $d = 1.001$, b) $d = 1.1$, c) $d=1.2$, d) $d = 1.3$. The solutions within the intervals $[-l_{cr}/2, l_{cr}/2]$ are shown with the red color.}
	\label{fig:FocusedSolitons}
\end{figure}

The waveforms of $N=20$ interacting solitons at the moment of focusing $t=0$ are shown in Fig.~\ref{fig:FocusedSolitons} for different values of the parameter $d$.
When $d$ is close to $1$, and all the soliton amplitudes in the ensemble are very close ($A_j \approx 1$, $j=1,...,N$), the effect of soliton repulsion becomes evident, see Fig.~\ref{fig:FocusedSolitons}a. It is clear that this pattern should be classified as an exchange type of interaction. 
Otherwise, when $d$ is sufficiently large, solitons focus into a one-humped wave with some fine oscillations of the shape (Fig.~\ref{fig:FocusedSolitons}d). According to \cite{Slunyaev2019}, the wave amplitude in the coordinate origin is exactly the alternating sum 
\begin{align}\label{Afocus}
	u_N(0,0) = A_1 - A_2 + A_3 - ... + (-1)^{N+1} A_N. 
\end{align}
This expression for the combination (\ref{geom}) can be explicitly calculated:
\begin{align}\label{AfocusCalculated}
	u_N(0,0) = \frac{d ( 1+(-1)^{N+1}d^{-N} )}{1+d} , \\
	u_N(0,0)  \underset{N \to \infty}{\longrightarrow}  \frac{d}{1+d}, \qquad
	u_N(0,0)  \underset{d \to 1+0}{\longrightarrow}  \left\{
	\begin{aligned}
		0, \quad \text{if } N \text{ is even} \\	1, \quad \text{if } N \text{ is odd.}
	\end{aligned} \right.  \nonumber
\end{align}
The value $u_N(0,0)$ in the limit of an infinite number of solitons with close amplitudes is ill-defined and depends on the order of taking limits. For a finite even number $N$ it is equal to zero, which is in good compliance with Fig.~\ref{fig:FocusedSolitons}a. On the contrary, another cases in Fig.~\ref{fig:FocusedSolitons}b-d correspond to the limit $N \to \infty$ with a fixed $d$.

% \begin{figure}[htp]
% 	{\includegraphics[width=7cm]{20s1.001.eps}}(a)
% 	{\includegraphics[width=7cm]{20s1.1.eps}}(b)
%	\\
%	{\includegraphics[width=7cm]{20s1.2.eps}}(c)
%	{\includegraphics[width=7cm]{20s1.3.eps}}(d)
%	\caption{20-soliton solution $u_{20}(x, t)$ at the moment of time $t = 0$ for %different values of $d$: a) $d = 1.001$, b) $d = 1.1$, c) $d=1.2$, d) $d = 1.3$.}
%	\label{fig:FocusedSolitons}
%\end{figure}

According to the examples in Fig.~\ref{fig:FocusedSolitons}, the characteristic value of $d$, which can approximately separate the exchange and overtaking types of interaction of a large number of solitons, should correspond to a remarkably smaller value than $3$, as it was in the case of two solitons. When the parameter $d$  increases above the value of about $1.2$, the characteristic waveform at the moment of the maximum focusing remains practically unchanged; then the value $u_N(0,0)$ characterises well the global maximum of the solution at $t=0$.

\begin{figure}[htp]
	\begin{centering}
		\includegraphics[width=7cm]{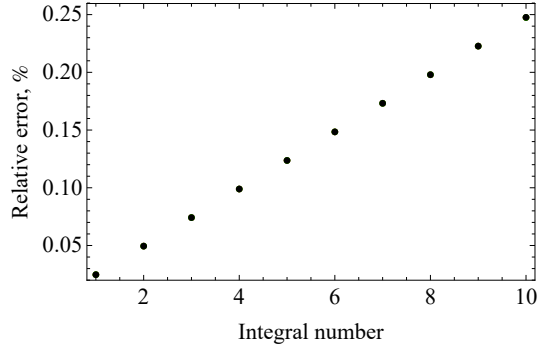}
		\caption{Relative errors of the first ten conservation integrals for the $20$-soliton solution with $d = 1.3$ at $t=0$.
			\label{fig:IntegralErrors}}
	\end{centering}
\end{figure}

The relative errors for the first $10$ conservation integrals $I_m$ (\ref{I_m}), which are compared to their analytical values (\ref{IntegralsNSoliton}), are given in Fig.~\ref{fig:IntegralErrors} for the case $N = 20$, $d = 1.3$ and the moment of time $t = 0$. The relative errors seem to grow linearly with an increase of the order of the integral. Note that the observed tendency to grow may characterise the accuracy of calculation of the integrals $I_m$ on the discrete grid, rather than the obtained solution $u(x,t)$. 
For various times $t$, and other values of the parameters $N$ and $d$ discussed in this paper, the reconstruction errors behave similarly to the case shown in Fig.~\ref{fig:IntegralErrors}. They are negligibly small.

\begin{figure}[htp]
	\begin{center}
		\includegraphics[width=16cm]{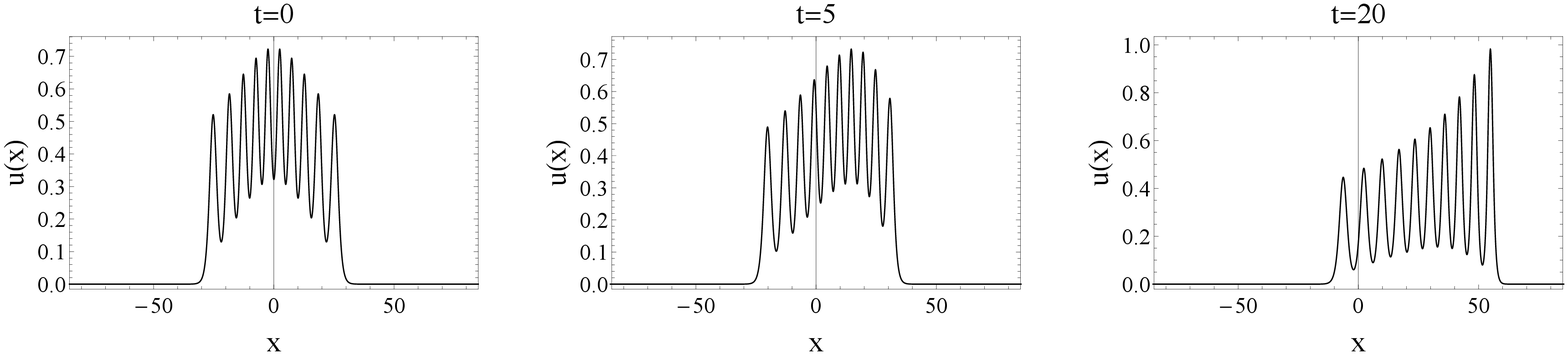}
		\caption{Interaction of 10 solitons, $d=1.1$.}
		\label{fig:Evolution1}
	\end{center}
\end{figure}

\begin{figure}[htp]
	\begin{center}
		\includegraphics[width=16cm]{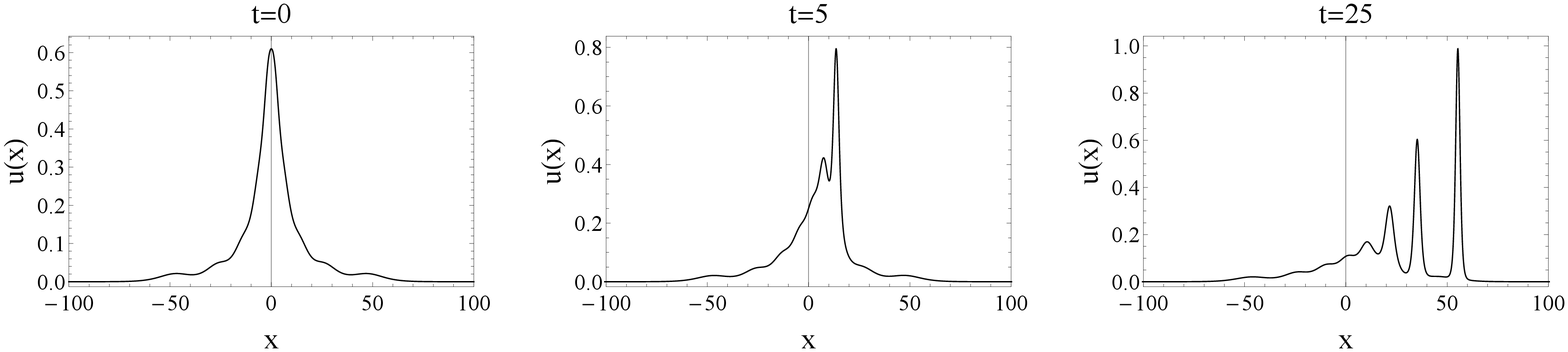}
		\caption{Interaction of 10 solitons, $d=1.6$.}
			\label{fig:Evolution2}
	\end{center}
\end{figure}

The evolution of $10$-soliton solutions for two different values of the parameter  $d$ is shown in Fig.~\ref{fig:Evolution1} and Fig.~\ref{fig:Evolution2}. Owing to the symmetry of the constructed solutions, it is enough to consider only non-negative moments of time $t \ge 0$. Initially, at $t = 0$, solitons are focused near the point $x = 0$. At large times solitons are separated in space in accordance to the dependence between their amplitudes and velocities. As one can conclude, the intermediate stage of an interaction of even a relatively small number of solitons can exhibit complicated dynamics. 

%\clearpage
\subsection{Statistical moments of the multisoliton fields}

In order to describe probabilistic characteristics of irregular waves, statistical moments are frequently used. In our work the statistical description is applied to particular realizations of a soliton gas, represented by exact $N$-soliton solutions, calculated as described in Sec.~\ref{sec:ExactSolutions}. We imply that the considered structure is a representative case of a simultaneous collision of $N$ solitons.
%with the given amplitude distribution (\ref{geom}).

For a given function $w(x)$ we introduce the space averaging in the certain interval $x \in [-l/2,l/2]$  as follows
\begin{align} \label{SpaceAveraging}
	\overline{w} = \frac{1}{l}\int_{-l/2}^{l/2}{w(x)dx}.
	%	\overline{w} = \lim_{l \rightarrow \infty}{\frac{1}{l}\int_{-l/2}^{l/2}{w(x)dx}}.
\end{align}
Assuming the homogeneity of $w(x)$, one can take the limit  $l \to \infty$ (as was done in \cite{el}), and then the value of $\overline{w}$ does not depend on the coordinate.
In our case the solution $u(x,t)$ tends to zero as $|x| \to \infty$; in the definition (\ref{SpaceAveraging}) we will assume that the considered area $[-l/2,l/2]$ corresponds to the domain of localization of the solution. Such a domain may be formally determined as the one for which the difference between the integrals $\int_{-l/2}^{l/2}{wdx}$ and $\int_{-\infty}^{\infty}{wdx}$ can be neglected. In what follows we will assume that this difference is completely negligible.

Thanks to the first two conserved integrals $I_1$ and $I_2$, the values $\overline{u}$ and  $\overline{u^2}$ can be calculated analytically at asymptotically large times, when the solitons constituting the solution overlap weakly \cite{Pelinovskyetal2013}:
\begin{align} \label{MeanValues}
	\overline{u} = 4 \rho \langle k \rangle, \qquad
	\overline{u^2} = \frac{16}{3} \rho \langle k^3 \rangle.
\end{align}
Here the averaging over the spectral parameter for $N$ solitons is introduced,
\begin{align}
	\langle \chi \rangle = \frac{1}{N} \sum_{j=1}^{N}{\chi_j},
\end{align}
and the parameter $\rho$ is used, which denotes the soliton gas density. This parameter is defined in an obvious way as the ratio of the number of solitons contained in the considered interval to the length of this interval,
\begin{align} \label{GasDensity}
	\rho = \frac{N}{l}.
\end{align}

With these definitions the formal expression for the variance can be converted to the following form:
\begin{align} \label{Variance}
	\sigma^2 = \overline{ ( u - \overline{u} ) ^2} = 
	\overline{u^2} - \overline{u}^2 = 
	\rho \frac{16}{3} \langle k^3 \rangle - 16 \rho^2 \langle k \rangle ^2.
	%	\sigma^2 = \overline{ ( u - \overline{u} ) ^2} = \overline{u^2} - \overline{u}^2 = \lim_{l \rightarrow \infty}{ \left( \rho \frac{16}{3} \langle k^3 \rangle - 16 \rho^2 \langle k \rangle ^2 \right)} .
\end{align}
In the situation of a rarefied gas, $\rho \ll 1$, the contribution from $\overline{u}$ becomes abnormally small, which leads to a simpler expression for the variance:
\begin{align}
	\sigma^2 \underset{\rho \ll 1}{\approx} \overline{u}^2 =  \frac{16}{3} \rho \langle k^3 \rangle.
\end{align}

In our situation, when the solution is strongly localized in space, the statistical moments $\mu_n$ introduced in the standard way depend on the size of the domain $l$, and for $n>2$ diverge when the size grows infinitely: 
\begin{align}
	\mu_n = \frac{\overline{\left( u - \overline{u} \right)^n}}{\sigma^n} \underset{\rho \ll 1}\approx l^{\frac{n}{2}-1} \frac{ \int_{-\infty}^{+\infty}{u^n dx}}{\left( \int_{-\infty}^{+\infty}{u^2 dx}  \right)^{\frac{n}{2}}}.
\end{align}
For this reason, further we will use a modified definition for the statistical moments as follows:
\begin{align}
	M_n (t) = \frac{1}{\Sigma^n} \int_{-\infty}^{+\infty} u^n(x,t) dx, \quad
	\Sigma^2 = \int_{-\infty}^{+\infty} u^2(x,t) dx, \\
	\mu_n  \underset{\rho \ll 1}\approx l^{\frac{n}{2}-1} M_n. \nonumber
\end{align}
They are independent on the choice of the consideration interval.

The first and the second moments, $M_1$, $M_2$, characterize the mean value of the field and the standard deviation respectively. They are related to the integrals $I_1$ and $I_2$ and hence do not change in time. Usually the third and the fourth statistical moments, which correspond to skewness and kurtosis respectively, are considered. They are not invariants of the KdV equation, and during the interaction of solitons the moments $M_3$ and $M_4$, calculated for $u_N(x,t)$, change their values.

At large times, when the solitons overlap negligibly weakly, the moments can be computed analytically. In terms of the spectral parameters the corresponding relations read \cite{Pelinovskyetal2013}:
\begin{align}
	M_1 = 4\frac{N \langle k \rangle}{\Sigma}
	, \quad \Sigma^2 = \frac{16}{3} N \langle k^3 \rangle,  \\
	M_2=1,\\
	M_3(\infty) = \lim_{t \rightarrow \pm \infty}{M_3(t)} = \frac{128}{15} \frac{N \langle k^5 \rangle}{\Sigma^3}, \label{M3} \\
	M_4(\infty) = \lim_{t \rightarrow \pm \infty}{M_4(t)} = \frac{512}{35} \frac{N \langle k^7 \rangle}{\Sigma^4} \label{M4}. 
\end{align}
For the power-law sequence of amplitudes (\ref{geom}) the expressions for the statistical moments can be computed explicitly using the formula for the sum of a geometric progression
\begin{align} \label{SumK}
	\sum_{j=1}^{N} k_j^n = 2^{-\frac{n}{2}} \frac{1-d^{-\frac{nN}{2}}}{1-d^{-\frac{n}{2}}}  \longrightarrow
	\left\{
	\begin{aligned}
		\frac{2^{-\frac{n}{2}}}{1-d^{-\frac{n}{2}}}, \quad  \text{if} \quad N \to \infty \\	
		2^{-\frac{n}{2}} N, \quad  \text{if} \quad d \to 1+0.
	\end{aligned} \right. 
\end{align}
When the number $N$ grows for a fixed $n$, the sums (\ref{SumK}) increase as well, but the statistical moments normalized to the variance, $M_3(\infty)$ and $M_4(\infty)$, decrease. We also note that the sums (\ref{SumK}) are bounded for all $d>1$, therefore the mean values $\langle k^n \rangle$ tend to zero for any power of $n $ as $N \rightarrow \infty$. This is a consequence of the particular choice of the distribution of soliton amplitudes.

\begin{figure}[htp]
	\begin{centering}
		\includegraphics[width=16cm]{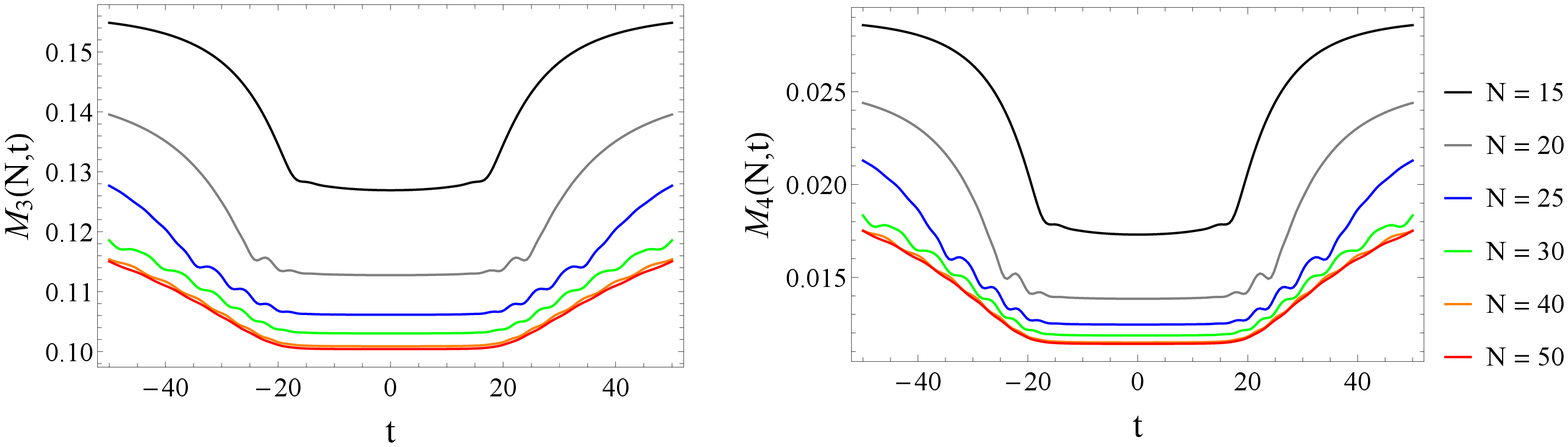}
		\caption{Evolution of the third and the fourth statistical moments for $N$-soliton solutions with $d = 1.1$ and $N=15...50$.} 
			\label{fig:Moments1}
	\end{centering}
\end{figure}

\begin{figure}[htp]
	\begin{centering}
		\includegraphics[width=16cm]{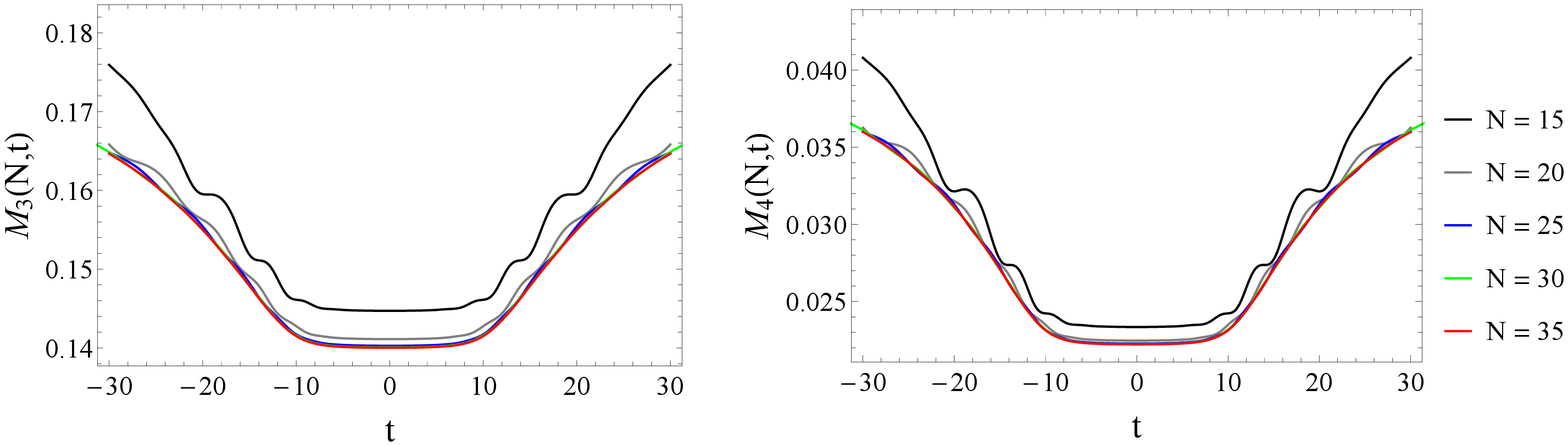}
		\caption{Evolution of the third and the fourth statistical moments for $N$-soliton solutions with $d = 1.2$ and  $N=15...35$.}
			\label{fig:Moments2}
	\end{centering}
\end{figure}

\begin{figure}[htp]
	\begin{centering}
		\includegraphics[width=16cm]{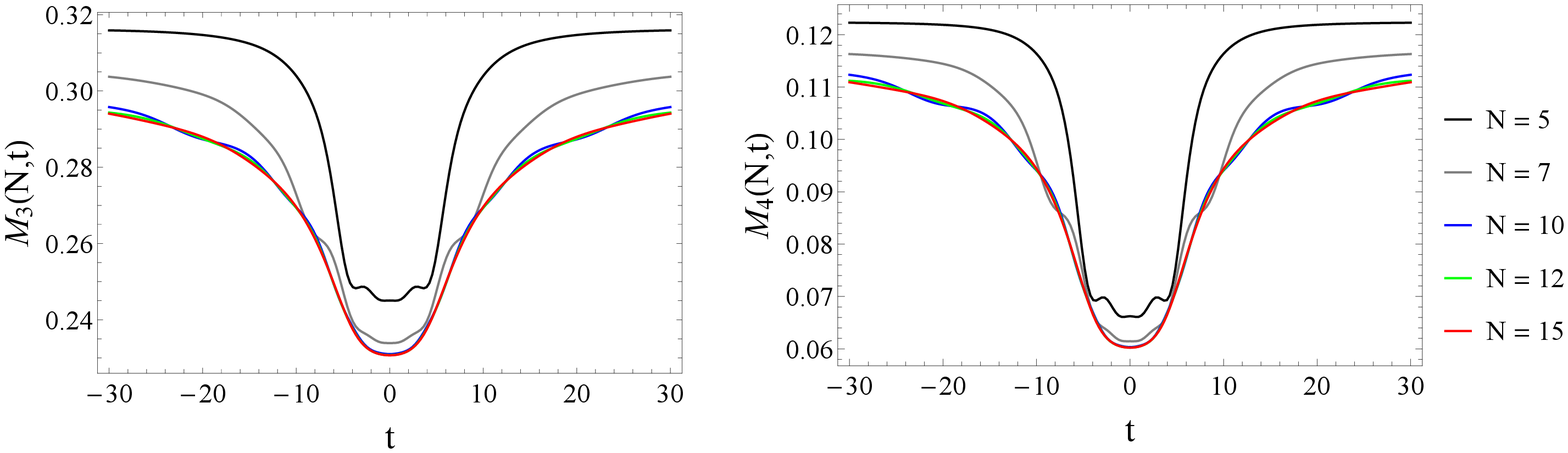}
		\caption{Evolution of the third and the fourth statistical moments for $N$-soliton solutions with $d = 1.6$ and  $N=5...15$.}
			\label{fig:Moments3}
	\end{centering}
\end{figure}

The evolution of the moments $M_3$ and $M_4$ in time for different number of solitons $N$ is shown in Fig.~\ref{fig:Moments1} -- \ref{fig:Moments3} for three different values of the parameter $d$. The dynamics of the third and the fourth moments are qualitatively similar.
As follows from the figures, the interaction of solitons of the Korteweg -- de Vries equation leads to a decrease of the moments compared to the values at asymptotically large times $M_3(\infty)$, $M_4(\infty)$, when the solitons are separated in space. An increase of the number $N$ results in a decrease of the moments for both, independent (at $t \to \pm \infty$) and focused (at $t \approx 0$) solitons. The graphs of dependences of the statistical moments on time exhibit long intervals in a vicinity of minima, which are characterized by just slight variations.
When the value of $d$ is very close to $1$, the plots of the moments demonstrate wide plateau areas, within which the values of the moments remain almost constant over time. Note that this observation is correct not only for exchange collisions (Fig.~\ref{fig:Moments1}), but also for overtaking ones (Fig.~\ref{fig:Moments2}), see Fig.~\ref{fig:FocusedSolitons}b and  Fig.~\ref{fig:FocusedSolitons}c 
respectively.

With an increase in $N$ the dependences $M_3(t)$ and $M_4(t)$ tend  to limiting ones for $N \to \infty$.
It follows from Figs.~\ref{fig:Moments1} -- \ref{fig:Moments3} that for larger values of the parameter $d$ the approaching to the limiting dependences occurs at smaller numbers $N$. One may conclude that accounting about 50 solitons is sufficient for a good approximation of these limiting dependences when $d \geq 1.1$. As $d$ increases, the extent of the ``plateau'' region reduces.

\begin{figure}[htp]
	{\includegraphics[width=7cm]{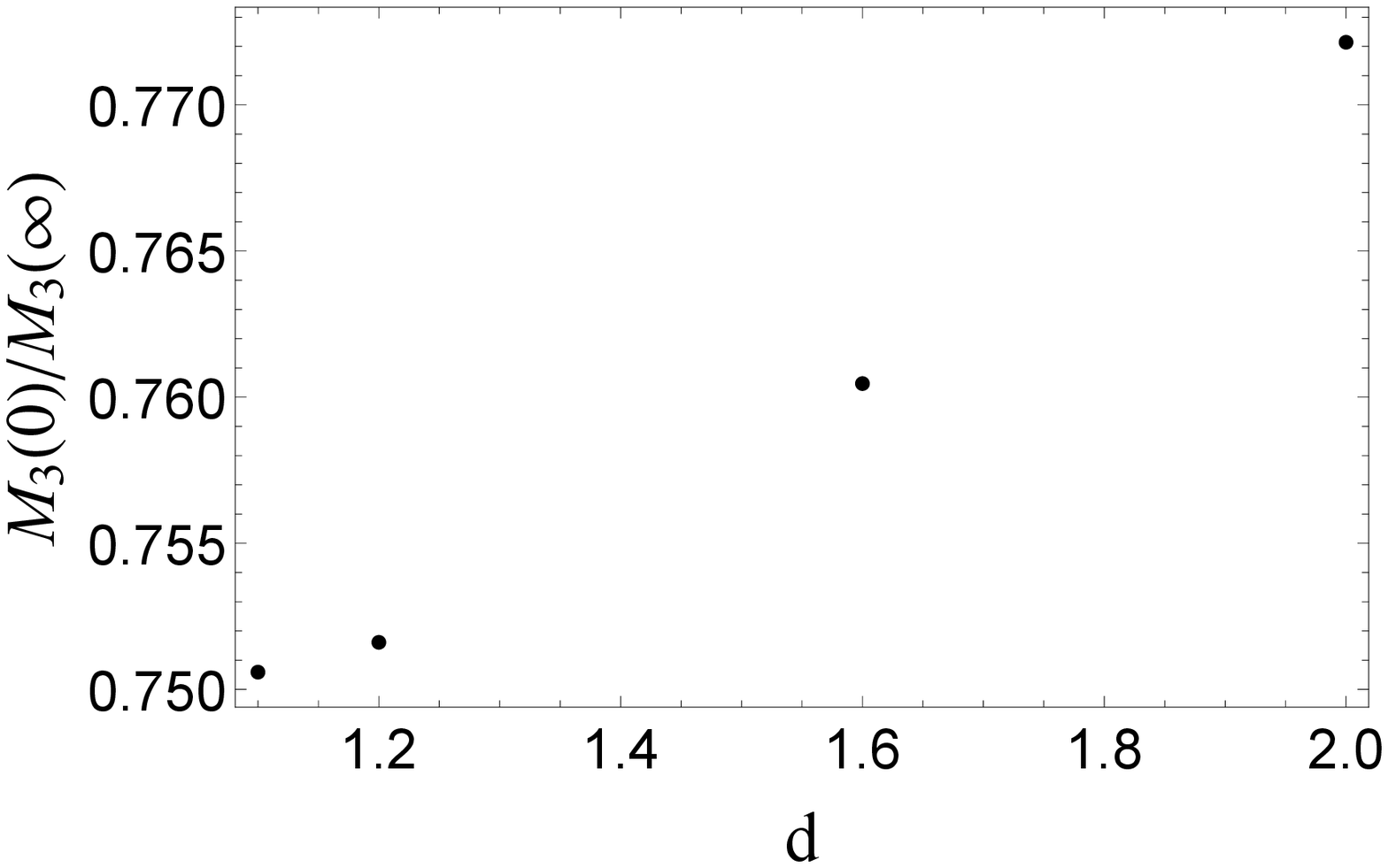}}(a)
	{\includegraphics[width=7cm]{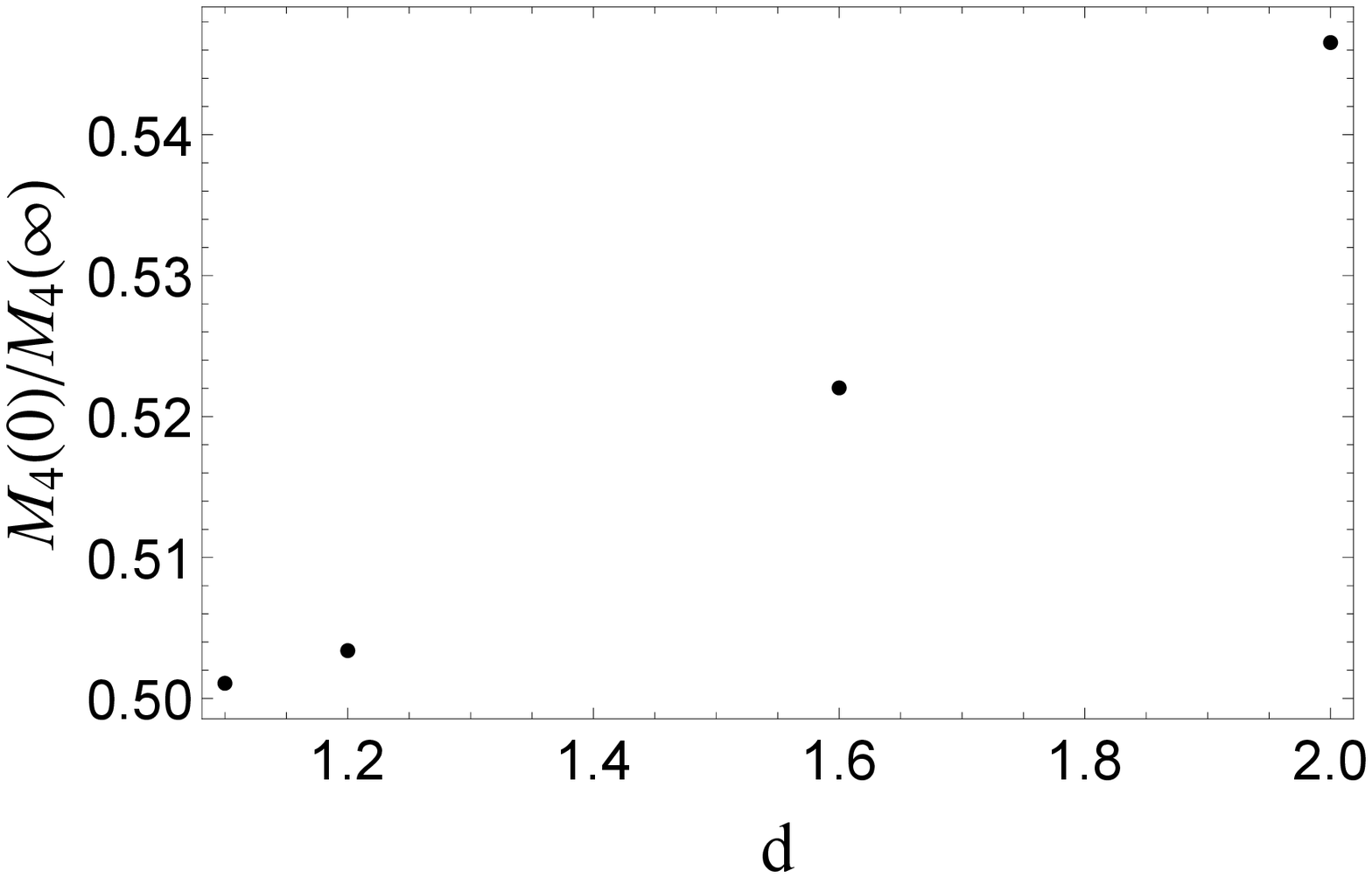}}(b)
	\caption{Relative reductions of the statistical moments $M_3(0)/M_3(\infty)$ (a) and $M_4(0)/M_4(\infty)$ (b) for $N=50$ solitons and different values of the parameter $d$.} 
		\label{fig:MomentsReduction}
\end{figure}

The graphs of the evolution of statistical moments similar to Figs.~\ref{fig:Moments1} -- \ref{fig:Moments3} were provided in \cite{Pelinovskyetal2013} for pairs of interacting KdV solitons. They demonstrated transient drops of the moments, though no pronounced ``plateaux'' could be found in those plots. 
The conclusion about the decrease in the statistical moments in many-soliton interactions is consistent with earlier results on the direct numerical simulation of ensembles of KdV solitons \cite{PelinovskyShurgalina2017} and the discussion of multisoliton collisions in the framework of the KdV equation in \cite{SlunyaevPelinovsky2016,Slunyaev2019}.

The values of skewness and kurtosis calculated numerically for the solution $u_N(x,0)$ at the instant of the maximum focusing are plotted in Fig.~\ref{fig:MomentsReduction}. In the graphs the moments are normalized to the values in the asymptotic state of independent solitons $|t|\rightarrow \infty$. In the limit $d \gg 1$ all solitons except the first one have negligibly small amplitudes, hence at large $d$ the dependences $M_3(0)/M_3(\infty)$ and $M_4(0)/M_4(\infty)$ should tend to $1$. It follows from the figure that during the soliton interaction the statistical moments reduce for any $d$. The most significant decrease corresponds to minimum values of the parameter $d$, when the sequence of soliton amplitudes decays slowly.

\subsection{Analytical estimates for the statistical moments.}

It is noteworthy that the dependences of statistical moments on time exhibit most extended ``plateau'' (Fig.~\ref{fig:Moments1}) with the smallest value of $d$. On this time interval the quantities $M_3$ and $M_4$ play the role of new (approximate) invariants of the solution, beyond the list of integrals $I_m$. This feature should obviously indicate in some sense degeneracy of the wave dynamics during the maximum focusing when $d$ is close to $1$.

Let us rewrite the integrals $I_3$ and $I_4$ for the equation (\ref{kdv}) in the following form:
\begin{align} \label{I3}
	I_3 &= 72 \left( a_1 - a_2 \right), \\
	a_1(t) &= \int_{-\infty}^{\infty}{u^3 dx}, \quad
	a_2(t) = \frac{1}{2} \int_{-\infty}^{\infty}{  (u_x)^2  dx}, \nonumber \\
	I_4 &= 324 \left( b_1 -b_2 + b_3\right), \label{I4}\\
	b_1(t) &=  \int_{-\infty}^{\infty}{u^4 dx}, \quad
	b_2(t)=  2\int_{-\infty}^{\infty}{u (u_x)^2 dx}, \quad
	b_3(t) = \frac{1}{5} \int_{-\infty}^{\infty}{ (u_{xx})^2 dx}. \nonumber
\end{align}
The $N$-soliton solution of the KdV equation can be presented in a sign-definite quadratic form \cite{Gardneretal1974}, then the solution at any instant of time is non-negative, $u_N(x,t) \ge 0$, $\forall (x,t)$. Therefore $a_1(t) > 0$ and $b_2 (t)>0$.
Then each of the integrals $a_1$, $a_2$, $b_1$, $b_2$, $b_3$ is non-negative. These integrals can be calculated analytically at the asymptotic stage $|t|\rightarrow \infty$:
\begin{align} \label{I3I4_Values}
	a_1(\infty) &= \frac{128}{15} N \langle k^5 \rangle, \quad
	a_2(\infty)=\frac{1}{4} a_1(\infty), \\
	b_1(\infty)  &= \frac{512}{35} N \langle k^7 \rangle, \quad
	b_2(\infty)=  \frac{2}{3} b_1(\infty), \quad
	b_3(\infty)= \frac{1}{6} b_1(\infty).
\end{align}
As was noted in \cite{Pelinovskyetal2013}, since the terms $a_1$ and $a_2$ enter the integral $I_3$ with different signs, the reduction of the integral $a_1(t) = \Sigma^3 M_3(t)$, responsible for asymmetry, is accompanied by a decrease in the second integral term $a_2(t)$, which contains the squared spatial derivative,
\begin{align} 
	\frac{M_3(t)}{M_3(\infty)} = \frac{a_1(t)}{a_1(\infty)} = \frac{3}{4} + \frac{a_2(t)}{a_1(\infty)},
\end{align}
which should correspond to some ``smoothing'' of the field. Since $a_2$ is always non-negative, the admissible absolute minimum of the integral $a_1$ (and of the statistical moment $M_3$) is reached when the value of $a_2$ tends to zero:
\begin{align} \label{MinimumAsymmetry}
	\frac{\min{M_3}}{M_3(\infty)} = \frac{3}{4} .
\end{align}
Note that the numerical solution with the parameters $d = 1.1$ and $N=50$ is characterised by a remarkably close to (\ref{MinimumAsymmetry}) value $M_3(0)/M_3(\infty) = 0.7506$, see Fig.~ \ref{fig:MomentsReduction}a and Table~\ref{tab:Moments}. Recall that according to Fig.~\ref{fig:Moments1}
this numerical solution seems to approximate well the limiting curve for the case of an infinite number $N$.

Assuming in a similar way that in the case of small $d>1$ the contributors with derivatives $b_2$ and $ b_3$ can be neglected in the integral $I_4$ at $t \approx 0$ as a consequence of the ``smoothing'' of the wave field in this time interval, we also set $ {b_1(t=0)} = b_1 \left( 1 - \frac{2}{3} + \frac{1}{6}\right)$, thus
\begin{align} \label{MinimumKurtosis}
	\frac{{M_4(t=0)}}{M_4(\infty)} = \frac{1}{2} .
\end{align}
The calculation based on the numerical solution almost coincides with this prediction, see Table~\ref{tab:Moments}.
We emphasize that the anomalous smallness of the integrals $a_2$, $b_2$, $b_3$ is not evident from the plots of the solution in Figs.~\ref{fig:FocusedSolitons}, \ref{fig:Evolution1}, \ref{fig:Evolution2}. Also note that the hypothetical minimal value of kurtosis, which follows from the form of the integral $I_4$ when choosing $b_2=0$ and $b_3 = \frac{1}{324}I_4$, is zero, $M_4=0$, which is unattainable by a solution of a finite amplitude.

We refer here to the paper \cite{Onoratoetal2016}, where the fact of conservation of the Hamiltonian of the nonlinear Schr\"odinger equation was used to relate the magnitude of the fourth statistical moment to the dimensionless spectrum width (the possibility of using the Hamiltonian of the KdV equation to relate the asymmetry parameter to the spectrum width was also noted there). Just as in the mentioned paper, we have not got an explanation of the physical mechanism of ``smoothing'' of the solution of the KdV equation in the course of the soliton focusing, but the fact of the smoothing becomes evident through the behavior of instantaneous statistical characteristics of the wave field. 

It is clear that the existence of an infinite number of conservation laws (\ref{I_m}) for the integrable equation (\ref{kdv}) allows one to obtain estimates even for higher statistical moments $M_n$, $n>4$. However, it can be expected that the growth of $n$ will be accompanied by a narrowing of the ``plateau'' intervals in the dependences of statistical moments and by a decrease in the accuracy of the analytical estimates.

We have tested this hypothesis by considering higher statistical moments. Table~\ref{tab:Moments} contains the values of the relative moments $M_n(0)/M_n(\infty)$ up to $n=7$. The analytical estimates based on the conservation integrals (obtained as shown above assuming that the integral parts with derivatives are negligibly small in integral sense) are given versus the results of numerical calculations based on the constructed $N$-soliton solutions $u_N(x,0)$. It follows from the table that the assumption of a wave field ``smoothing'' at the stage of the maximum focusing provides excellent estimates for all considered high-order statistical moments. The interaction of solitons of the KdV equation results in decreasing of not only the moments $M_3$ and $M_4$, but also of the moments of higher orders. Moreover, a degree of reduction of a statistical moment value increases with the sequence number.

\begin{table}[h]
	\caption{The minimum relative values of the $n$-th statistical moments $M_n(0)/M_n(\infty)$. The numerical solution corresponds to the case $d=1.1$ and $N=50$.}
	\centering
	\begin{tabular}{ | c | c | c | c |} 	\hline
		$n$ & analytical estimate &  numerical solution & difference \\ \hline
		$3$ & $3/4$ & 0.7506 & 0.08\% \\
		$4$ & $1/2$ & 0.5011 & 0.22\% \\
		$5$ & $5/16$  & 0.3138 & 0.42\% \\
		$6$ & $3/16$ & 0.1889 & 0.72\% \\
		$7$ & $7/64$ & 0.1106 & 1.1\% \\
		\hline
	\end{tabular}
	\label{tab:Moments}
\end{table}

\subsection{The state of critical density} \label{sec:CriticalDensity}

Since the number $N$ is conserved during the evolution of the solution $u_N(x,t)$, the soliton density estimate (\ref{GasDensity}) depends only on the characteristic size $l$ of the region where the solution is localized. It evidently becomes infinitely large at $t \to \pm \infty$ (then the density $\rho$ tends to zero) and decreases when solitons collide.
The formal requirement that the variance $\sigma^2$ (\ref{Variance}) may not be negative leads to the condition on the maximum allowed density of the soliton, $\rho_{cr}$, when $\overline{u}^2 = \overline{u^2}$ \cite{PelinovskyShurgalina2017, el}
\begin{align} \label{CriticalDensity}
	\rho_{cr} = \frac{\langle k^3 \rangle}{3 \langle k \rangle^2}.
\end{align}

It is quite natural to relate the time interval of strong compression of solitons, accompanied by ``frozen'' in time statistical moments, 
with the state of the maximum (critical) density of solitons.
%with the situation of \red{the} realization of the maximum (critical) density of solitons.
%
For the considered soliton amplitude distribution (\ref{geom}) the critical density (\ref{CriticalDensity}) is calculated with the use of (\ref{SumK}):
\begin{align} \label{CriticalDensityValue}
	\rho_{cr}  \underset{N \gg 1}{\longrightarrow} \frac{N}{3 \sqrt{2}} \frac{(1-d^{-\frac{1}{2}})^2}{1-d^{-\frac{3}{2}}}.
\end{align}
It follows from this formula that the critical density for a large number of solitons $N$ is minimal when $d$ approaches 1, and gradually increases with $d$ up to the value $N/(3\sqrt{2})$ when $d$ tends to infinity. For a fixed $d$, the critical density is proportional to $N$.

\begin{figure}[h]
	\begin{centering}
		\includegraphics[width=9cm]{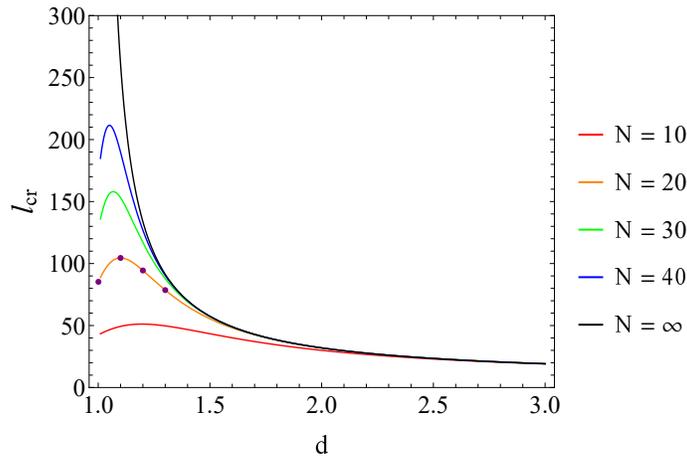} 
		\caption{The critical length $l_{cr}$ as a function of the parameter $d$ for different values of $N$. The full circles correspond to the examples shown in Fig.~\ref{fig:FocusedSolitons}.}
		\label{fig:CriticalLength}
	\end{centering}
\end{figure}

According to the definition of the density (\ref{GasDensity}), for a fixed number of solitons $N$ the critical density corresponds to the minimal effective region of localization of the soliton field, which can be characterized by the size $l_{cr}=N/\rho_{cr}$. As follows from (\ref{CriticalDensityValue}), for a large number of solitons with amplitudes distributed according to a power law, the value $l_{cr}$ depends only on the parameter of the amplitude distribution $d$. In the limit of large $d$ the width $l_{cr}$ tends to the value $l_{sol}=3\sqrt{2}$, which can be considered as an integral width of one soliton of unit amplitude. When the parameter $d$ tends to 1 having infinite $N$, the value $l_{cr}$ increases indefinitely, $l_{cr} \approx 18\sqrt{2}(d-1)^{-1}$, as shown in Fig.~\ref{fig:CriticalLength} with the black curve. 

Note that when $d \rightarrow 1+0$, but $N$ remains finite, the sums of the spectral parameters in (\ref{CriticalDensity}) have limiting values which differ from the case $N \rightarrow \infty$. Then the critical density is finite, 
\begin{align} \label{CriticalDensityValue2}
	\rho_{cr}  \underset{d \to 1+0}{\longrightarrow} \frac{1}{3 \sqrt{2}},
\end{align}
and $l_{cr} \underset{d \to 1+0}{\longrightarrow} N l_{sol}$. In other words, a finite number of solitons with very close amplitudes effectively do not compress upon focusing.  
The functions $l_{cr} = N/\rho_{cr}$ calculated for several choices of finite numbers $N$ are shown in Fig.~\ref{fig:CriticalLength} with color curves. According to the plots, these dependences are not monotonic in the range of relatively small $d$.

The values of $l_{cr}$ calculated for the examples shown in  Fig.~\ref{fig:FocusedSolitons} ($N=20$) are represented in Fig.~\ref{fig:CriticalLength} by full circles. The corresponding intervals $[-l_{cr},l_{cr}]$ are shown in Fig.~\ref{fig:FocusedSolitons} by parts of the solutions highlighted in red. One may conclude that the values of $l_{cr}$ give reasonable estimates for the focused wave sizes. For instance, the critical width for $d = 1.001$ (Fig.~\ref{fig:FocusedSolitons}a) is $l_{cr} \approx 85$, which is about two time less than the actual visible area occupied by the solution $u_{20}(x,0)$.

%		\includegraphics[width=7cm]{l_cr_points.eps} 
%		\caption{The critical length $l_{cr}$ as a function of the parameter $d$ in the limit of a large number of solitons $N \gg 1$, and the values $l_{cr}$, calculated for a finite value $N = 20$ when $d = 1.001$, $d = 1.1$, $d = 1.2$, $d = 1.3$.}
%\label{fig:CriticalLength}
%	\end{centering}
%\end{figure}

%\begin{figure}[htp]
%	\begin{centering}
%		\includegraphics[width=7cm]{l_cr_new.eps} 
%		\caption{The critical length $l_{cr}$ as a function of the parameter $d$ in the limit of a large number of solitons $N \gg 1$.}
%			\label{fig:CriticalLength}
%	\end{centering}
%\end{figure}

\section{Conclusions} \label{sec:Conclusion}

We further develop the suggested in \cite{Pelinovskyetal2013}  approach to study the complicated dynamics and statistical features of a soliton gas (i.e. large ensembles of solitons) via examination of toy problems that are representative but still admit analytic or semi-analytic solution.
In \cite{Pelinovskyetal2013} the simplest case of the interaction of two solitons was considered. In the present work we have analysed exact multisoliton solutions of the Korteweg -- de Vries equation, including the limit of an infinitely large ensemble with a finite energy (in the sense of the integral of the squared solution).

Our research is focused on the synchronous soliton collisions, which should correspond to the strongest focusing of solitons. The events of simultaneous collisions of a large number of solitons and/or breathers which occurred in direct numerical simulations were noted in  \cite{GelashAgafontsev2018,Didenkulova2019}. They were investigated analytically in the framework of various integrable equations with a focus on the emergence of abnormally high waves \cite{Akhmedievetal2009,Chinetal2016,SlunyaevPelinovsky2016,Sun2016,Slunyaev2019}. These studies have shown that multiple collisions of solitons of the KdV equation do not cause waves of higher amplitude.
The density of the soliton gas is bounded above by the value of the critical density, which was obtained on formal grounds in \cite{PelinovskyShurgalina2017} and more systematically in \cite{el}.

To study the properties of soliton fields, we have implemented a numerical procedure for constructing $N$-soliton exact solutions of the KdV equation using the Darboux transformation. The procedure made it possible to calculate solutions with a very high accuracy up to the order of $N=50$.
In a recent preprint \cite{Bonnemainetal2022} the Darboux transformation was used to construct $N$-soliton solutions of the KdV equation with $N$ up to an even larger number of the order of $200$. It was noticed there that for an accurate construction of the $N$-soliton solution it was necessary to set the length of the mantissa to $2N$ signs. In our paper $100$-digits arithmetic is used to calculate at most $50$ solitons, which agrees well with that estimate. Due to rapidly growing computational costs incurred during a construction of solutions with large $N$, the most efficient strategy for studying a gas of solitons with random parameters seems to be an accumulation of statistics by averaging over an ensemble of realizations consisting of a relatively small number of solitons.

In the present paper the distribution of soliton amplitudes in accordance with a geometrical progression with the common ratio $d^{-1}$ was considered as a model one which ensures finite integral characteristics of the ensemble even in the limit $N \to \infty$.
Depending on the value of $d$, the interactions of a large number of solitons look similar to the ``overtaking'' or ``exchange'' scenarios of interactions of pairs of solitons, although these regimes can be differentiated less clearly than in the case of two solitons, and at a smaller value of $d>1$.

It is shown that simultaneous collisions of any number of solitons lead to a decrease of the statistical moments of asymmetry and kurtosis, as well as the moments of higher orders (up to the seventh moment have been checked).
When the number of solitons $N$ grows, the curves of the evolution of the statistical moments converge to the limiting ones for $N \to \infty$. The limiting curves can be well approximated with the help of the solution for a finite number $N$ for the values of $d>1$ which are not too close to 1.

During the stage of maximum focusing of a large number of KdV solitons with significant amplitudes (the parameter $d>1$ is relatively small), long time intervals of a quasi-stationary behavior of the statistical moments appear, which effectively correspond to new approximate conservation integrals. It is shown that these time intervals are characterized by ``smoothed'' fields with derivatives which are small in integral sense.
With the use of this property, analytical expressions for the reduced values of high-order statistical moments are obtained, which perfectly agree with the numerical solution. In particular, it is found that for the third statistical moment (asymmetry) the minimum possible value is reached at the stage of the maximum focusing.

The intervals of quasi-stationarity of the statistical moments, which correspond to the maximum compression of soliton fields, are associated with situations when the critical (maximum) density of the soliton gas is realized. The analytical estimate of the minimum size of the region occupied by the soliton field, which follows from this assumption, is in reasonable agreement with the exact solution.

\vspace{0.25cm}

{\bf Acknowledgements.} The study of the critical density state was supported by Laboratory of Dynamical Systems and Applications NRU HSE, of the Ministry of Science and Higher Education of the RF grant Ag. No. 075-15-2019-1931. The remaining research was supported by the RSF grant No. 19-12-00253. The authors are grateful to E.N. Pelinovsky for his valuable comments.

\appendix
\section{First 10 conservation laws for the KdV equation in explicit form} \label{sec:Appendix}

The conservation laws are reproduced in the form used in \cite{miura}, for the scaled function $Q(x,t)$ and its $n$-th derivatives with respect to the coordinate $x$ denoted as $Q_n$: 
\begin{align}
	Q(x,t) = 6u(x,t), \quad  Q_n = \frac{\partial^n}{\partial x^n}Q,  \quad Q_0 = Q; 
\end{align}
\begin{align} 
	T_1 = Q_0,
\end{align}
\begin{align}
	T_2 = \frac{1}{2} Q_0^{2},
\end{align}
\begin{align}
	T_3 = \frac{1}{3} Q_0^{3} - Q_1^{2},
\end{align}
\begin{align}
	T_4 = \frac{1}{4} Q_0^{4} - 3Q_0 Q_1^{2} + \frac{9}{5}Q_2^{2},
\end{align}
\begin{align}
	T_5 = \frac{1}{5} Q_0^{5} - 6Q_0^2 Q_1^{2} + \frac{36}{5}Q_0 Q_2^{2} - \frac{108}{35}Q_3^2, 
\end{align}
\begin{align}
	T_6 = \frac{1}{6} Q_0^{6} - 10Q_0^3Q_1^2 + 18Q_0^2Q_2^2 - 5Q_1^4 - \frac{108}{7}Q_0 Q_3^2 + \frac{120}{7}Q_2^3 + \frac{36}{7}Q_4^2,
\end{align}
\begin{align} 
	T_7 = \frac{1}{7} Q_0^7 - 15 Q_0^4 Q_1^2 + 36 Q_0^3 Q_2^2 - 30 Q_0 Q_1^4 - 
	\frac{324}{7} Q_0^2 Q_3^2 + \frac{720}{7} Q_0 Q_2^3 + 108 Q_1^2 Q_2^2 \nonumber \\ 	
	+ \frac{216}{7} Q_0 Q_4^2 - \frac{1080}{7} Q_2 Q_3^2 - \frac{648}{77} Q_5^2,
\end{align}
\begin{align} 
	T_8 = \frac{1}{8} Q_0^8 - 21 Q_0^5 Q_1^2 + 63 Q_0^4 Q_2^2 - 105 Q_0^2 Q_1^4 - 
	108 Q_0^3 Q_3^2 + 360 Q_0^2 Q_2^3  \nonumber \\
	+ 756 Q_0 Q_1^2 Q_2^2 + 108 Q_0^2 Q_4^2-324 Q_1^2 Q_3^2 - 1080 Q_0 Q_2 Q_3^2 + 378 Q_2^4 - \frac{648}{11} Q_0 Q_5^2 \nonumber \\
	 + 	\frac{4536}{11} Q_2 Q_4^2 + \frac{1944}{143} Q_6^2,
\end{align}
\begin{align} 
	T_9 = \frac{1}{9} Q_0^9 - 28 Q_0^6 Q_1^2 + \frac{504}{5} Q_0^5 Q_2^2 - 280 Q_0^3 Q_1^4 - 
	216 Q_0^4 Q_3^2 + 960 Q_0^3 Q_2^3 \nonumber \\
	+ 3024 Q_0^2 Q_1^2 Q_2^2 - 168 Q_1^6 + 288 Q_0^3 Q_4^2 - 4320 Q_0^2 Q_2 Q_3^2 - 2592 Q_0 Q_1^2 Q_3^2  \nonumber \\
	+ 
	3024 Q_0 Q_2^4 + \frac{26496}{5} Q_1^2 Q_2^3 - \frac{2592}{11} Q_0^2 Q_5^2 + \frac{36288}{11} Q_0 Q_2 Q_4^2 +
	864 Q_1^2 Q_4^2  \nonumber  \\
	- \frac{169344}{55} Q_1 Q_3^3 - \frac{545184}{55} Q_2^2 Q_3^2 + \frac{15552}{143} Q_0 Q_6^2 - \frac{145152}{143} Q_2 Q_5^2  \nonumber \\
	 + 	 \frac{653184}{715} Q_4^3 - \frac{15552}{715} Q_7^2,
\end{align}
\begin{align} 
	T_{10} = \frac{1}{10} Q_0^{10} - 36 Q_0^7 Q_1^2 - 630 Q_0^4 Q_1^4 + \frac{756}{5} Q_0^6 Q_2^2 - 1512 Q_0 Q_1^6 + 2160 Q_0^4 Q_2^3 \nonumber \\
	+ 9072 Q_0^3 Q_1^2 Q_2^2 - \frac{1944}{5} Q_0^5 Q_3^2 + 13608 Q_0^2 Q_2^4 + \frac{238464}{5} Q_0 Q_1^2 Q_2^3  \nonumber \\
	+ 13608 Q_1^4 Q_2^2 - 12960 Q_0^3 Q_2 Q_3^2 - 11664 Q_0^2 Q_1^2 Q_3^2 + 648 Q_0^4 Q_4^2
	+ \frac{178848}{11} Q_2^5  \nonumber \\
	- \frac{1524096}{55} Q_0 Q_1 Q_3^3 - 
	\frac{4906656}{55} Q_0 Q_2^2 Q_3^2  
	- \frac{334368}{5} Q_1^2 Q_2 Q_3^2 + \frac{163296}{11} Q_0^2 Q_2 Q_4^2 \nonumber \\
	 + 7776 Q_0 Q_1^2 Q_4^2 - \frac{7776}{11} Q_0^3 Q_5^2 - 
	\frac{11045808}{715} Q_3^4   
	 + \frac{5878656}{715} Q_0 Q_4^3 \nonumber \\
	 + \frac{22208256}{715} Q_1 Q_3 Q_4^2 
	 + \frac{26570592}{715} Q_2^2 Q_4^2 - \frac{1306368}{143} Q_0 Q_2 Q_5^2 - 
	\frac{23328}{11} Q_1^2 Q_5^2   \nonumber \\
	+ \frac{69984}{143} Q_0^2 Q_6^2 - \frac{5878656}{715} Q_4 Q_5^2 + 
	\frac{1679616}{715} Q_2 Q_6^2 - \frac{139968}{715} Q_0 Q_7^2 + \frac{419904}{12155} Q_8^2.
\end{align}

\end{document}